%% file: Optimal-Position-Managment-rv.tex
\documentclass[a4,11pt]{article}
\usepackage{graphicx} 
\usepackage{float}
\usepackage{amsmath,amsthm}
\usepackage{amsfonts}
\usepackage{latexsym}
\usepackage{amssymb}
\usepackage{setspace}
\usepackage{comment}

\makeatletter
 
  \@addtoreset{equation}{section}
 \makeatother

\begin{document}
\title{Optimal Position Management for a Market Maker\\ with
Stochastic Price Impacts~\footnote{
All the contents expressed in this research are solely those of the author and do not represent any views or 
opinions of any institutions. The author is not responsible or liable in any manner for any losses and/or damages caused by the use of any contents in this research.
}
}

\author{Masaaki Fujii\footnote{Graduate School of Economics, The University of Tokyo. e-mail: mfujii@e.u-tokyo.ac.jp}
}
\date{ \small 
5 September, 2015
}
\maketitle


\input{tymacro3.TEX}


\def\E{{\bf E}}
\def\P{{\bf P}}
\def\Q{{\bf Q}}
\def\R{{\bf R}}

\def\cadlag{{c\`adl\`ag~}}

\def\calf{{\cal F}}
\def\calp{{\cal P}}
\def\calq{{\cal Q}}
\def\wtW{\widetilde{W}}
\def\wtB{\widetilde{B}}
\def\wtPsi{\widetilde{\Psi}}
\def\wt{\widetilde}
\def\mbb{\mathbb}
\def\opi{{\pi^*}}
\def\odel{{\del^*}}
\def\pii{{\pi^i}}
\def\pidel{{\pi,\del}}
\newcommand{\bvec}[1]{\mbox{\boldmath $#1$}}
\def\bpi{\bvec{\pi}}
\def\bX{\bvec{X}}
\def\bx{\bvec{x}}
\def\by{\bvec{y}}
\def\bee{\bvec{e}}
\def\bbs{\bvec{b}}
\def\bS{\bvec{S}}
\def\bdel{\bvec{\del}}
\def\bPhi{\bvec{\Phi}}
\def\bPsi{\bvec{\Psi}}
\def\bbeta{\bvec{\beta}}
\def\bl{\bvec{l}}
\def\bdel{\bvec{\del}}
\def\bTheta{\bvec{\Theta}}
\def\btheta{\bvec{\theta}}

\def\LDis{\frac{\bigl.}{\bigr.}}
\def\ep{\epsilon}
\def\del{\delta}
\def\part{\partial}
\def\wh{\widehat}
\def\bsigma{\bar{\sigma}}
\def\yy{\\ && \nonumber \\}
\def\y{\vspace*{3mm}\\}
\def\nn{\nonumber}
\def\be{\begin{equation}}
\def\ee{\end{equation}}
\def\bea{\begin{eqnarray}}
\def\eea{\end{eqnarray}}
\def\beas{\begin{eqnarray*}}
\def\eeas{\end{eqnarray*}}
\def\l{\left}
\def\r{\right}

\def\bull{$\bullet~$}

\newcommand{\Slash}[1]{{\ooalign{\hfil/\hfil\crcr$#1$}}}

\begin{abstract}
This paper deals with an optimal position management problem for a market maker
who has to face uncertain customer order flows in an {\it illiquid} market,
where the market maker's continuous trading incurs
a {\it stochastic} linear price impact.
Although the execution timing is uncertain, 
the market maker can also ask its OTC counterparties to transact a block trade without causing 
a direct price impact.
We adopt quite generic stochastic processes of the securities, order flows,
price impacts, quadratic penalties as well as security borrowing/lending rates.
The solution of the market maker's optimal position-management strategy is represented by a stochastic Hamilton-Jacobi-Bellman equation,
which can be decomposed into three (one non-linear and two linear) backward stochastic differential equations (BSDEs).
We provide the verification using the standard BSDE techniques for a single security case.
For a multiple-security case, we make use of the connection of the non-linear BSDE 
to a special type of backward stochastic Riccati differential equation (BSRDE)
whose properties were studied by Bismut~(1976).
We also propose a perturbative approximation scheme  
for the resultant BSRDE,  which only requires a system of linear ODEs to be solved
at each expansion order. Its justification and the convergence rate are also given.

\end{abstract}
\vspace{0mm}

\section{Introduction}
The financial market currently being formed in the aftermath of the great financial crisis 
looks completely different from the previous one. 
Mandatory clearing for the standardized financial products and much higher regulatory costs for 
the rest of over-the-counter (OTC) contracts made many investors withdraw from 
the long-dated exotic derivative business and pay much attention to the trading of 
listed products with exchanges or standard contracts with central counterparties.

In the new market, it is clear that the exchanges and central counterparties are the most
important trading venues and have started to play a much bigger role than before.
However, these new developments have not completely diminished the importance of 
the traditional key players in the market, that is a {\it market maker}.
A market maker is a firm that quotes buy and sell prices for financial securities and derivatives,
and stands ready to perform these deals on a regular and/or continuous basis.
They are crucially important to maintain liquidity for equities, currencies, commodities, government/corporate bonds,
many structured products and derivatives.
Even for products tradable at an exchange, 
market makers are playing an important role by intermediating
non-financial corporates and other investors
since it is not always possible for them to satisfy  many regulatory conditions required to 
get a direct access to the exchange.
There exist many other benefits such as those related to accounting, anonymity and flexibility  that may be
obtained with the help of market makers.

Especially due to the proposed regulation on the leverage ratio and the higher capital amount
required for the open positions,  the market makers have to deal with formidable tasks. Due to the 
smaller warehousing capability of their balance sheets, they need more active 
position management. At the same time, they have to optimize 
execution strategies in order to avoid unnecessarily big market impacts
and the associated transaction costs. See \cite{Risk-net}, for example, and other articles
available in {\it Risk.net} to get some images of the recent market.

In this paper, we consider the optimal position management problem for a market maker who is 
facing uncertain customer orders. We are interested in a {\it good} market maker who
accepts every customer order with a predefined  bid/offer spread. The spread
can be stochastic but we do not allow the market maker to control its size
dynamically based on its proprietary reasons in order to give a bias to the 
customer flows. Otherwise the firm will not be considered as a trustful 
market maker~\footnote{In fact, this was the common pride I observed among the fellow traders while I was in
the industry.}.
We suppose that there exists a relatively liquid market for security borrowing and lending (i.e., 
so called repo transactions), which can be used by the market maker to answer the incoming customer orders.
In addition to  matching an incoming order directly to the security being warehoused in its balance sheet,
it is assumed that the market maker can access two external trading venues.
One is a traditional exchange where the market maker carries out absolutely continuous trading
that incurs, however, a stochastic linear price impact. 
It is also supposed that
the participants of the exchange can {\it partially} infer the inventory size of 
the market maker and that their aggregate reactions, as a preparation for the market maker's future unwinding,
affect proportionally to the security price. 
Another venue is the aggregate of the market maker's
OTC counterparties with which the firm can execute a block trade without 
directly affecting the price in the exchange. In this case, however,  the execution timing is  uncertain.
The modeling of the latter venue (we call it the dark pool) is closely related to the one introduced by 
Kratz \& Sch\"oneborn (2013)~\cite{Kratz} except that we allow stochasticity in its execution intensity. 

There now exist the vast literature on the optimal execution problems.
Our model is closely related to the line of developments made by
Bertsimas \& Lo (1998)~\cite{Bertsimas}, 
Almgren \& Chriss (1999, 2000)~\cite{Almgren1,Almgren2}, 
Schied \& Sch\"oneborn (2009)~\cite{Schied-2009} and 
to more recent works Ankirchner \& Kruse (2013)~\cite{Ankirchner-Kruse}, Ankirchner, Jeanblanc \& Kruse~(2014)~\cite{Jeanblanc}
and Kratz \& Sch\"oneborn (2013)~\cite{Kratz}. In this first approach, the price process of the relevant security 
is exogenously modeled.
There exist many other interesting approaches, such as models of supply curves~( See, for example,  Bank \& Baum~(2004)~\cite{Bank}, Cetin, Jarrow \& Protter~(2004)~\cite{Cetin} and Roch (2011)~\cite{Roch}. ) and those directly modeling 
the dynamics of Limit Order Books~( See, for example, Obizhaeva \& Wang (2013)~\cite{Obizhaeva},
Alfonsi, Fruth \& Schied~(2008, 2010)~\cite{Alfonsi-2008, Alfonsi-2010}, Fruth \& Sch\"oneborn (2014)~\cite{Fruth}, and Cartea \& Jaimungal~(2015a,b)~\cite{Cartea-1,Cartea-2}. ).
We refer to the review articles Gatheral \& Schied (2013)~\cite{Gatheral-review}
and G\"okay, Roch \& Soner (2011)~\cite{Gokay} for the recent developments,
various other aspects and references.

The two main differences of the current work are the focus on the market maker's position management 
problem with uncertain customer orders, and the generality in its setup.
We allow quite generic stochastic processes for the securities,  position impact factors, compensators of incoming customer 
orders, execution intensities of the dark pools,  repo rates relevant for 
the security borrowing/lending, and the quadratic penalties for the outstanding position size etc. 
To the best of our knowledge, it is the most generic setup among the literature 
adopting the first approach in the last paragraph.
The resultant optimal strategy becomes fully adapted to the market filtration 
instead of a deterministic one usually found in the literature.
In contrast to a very short term liquidation strategy for which non-random (or even constant) coefficients
may be sufficient, this generality is necessary for the position management problem 
with a longer time horizon in which a significant change of the market conditions is naturally expected.
Although the first approach we choose is a simplified reduced-form approximation of Limit Order Books,
it allows more flexible modeling of the underlying processes including multiple securities and also their
mutual dependence, which is expected to be more relevant for our medium term  problem.

We follow the technique proposed by Mania \& Tevzadze (2003)~\cite{Mania} to derive 
the relevant stochastic HJB equation and its decomposition into three (one non-linear and two linear)
backward stochastic differential equations (BSDEs).
For a single security case, we use the standard results on BSDEs and the comparison theorem
(See, for example, Ma \& Yong (2007)~\cite{Ma} and Pardoux \& Rascanu (2014)~\cite{Pardoux}.) 
for verifying the solution. For a multiple-security case, however, we need to handle a
matrix-valued non-linear BSDE for which we do not have an appropriate comparison theorem.
We show that the relevant BSDE is actually a special type of 
backward stochastic Riccati differential equations (BSRDEs)
associated with a stochastic linear quadratic control (SLQC) problem.
Interestingly, a seemingly quite different setup of the optimization problem gives rise to the 
same BSDE. Thanks to this relation, we can guarantee the existence of a uniformly bounded solution 
by theorems proved by Bismut (1976)~\cite{Bismut}.
The main difficulty for the implementation of the proposed scheme is 
the concrete evaluation of this BSRDE.
We propose a perturbative expansion technique for the BSRDE with a general 
Markovian factor process,
which only requires to solve a system of linear ODEs at each order of expansion.
A justification and convergence rate of the approximation scheme are also given.

The organization of the paper is as follows: Sections 2 and 3 give some preliminaries,
the detailed market description and the market maker's problem.
Section 4 gives the  derivation of the candidate solution and its verification.
An extension to a multiple-security case is 
given in Sections 5 and 6.  Sections 7 and 8 deal with implementation. 
In particular, the perturbative scheme and its error estimate are given in Section 8. 
The behavior of the terminal position size with respect to the penalty size is 
studied in Appendix A.

\section{Preliminaries}
We consider a complete filtered probability space, in which all the stochastic processes are defined,
$(\Omega,\calf,\mbb{F},\mbb{P})$
where $\mbb{F}=(\calf_t)_{t\geq 0}$ is the filtration satisfying 
the usual conditions. $W$ is the $d$-dimensional standard Brownian motion 
and the $\mbb{P}$-augmented filtration generated by $W$ is denoted by $\mbb{F}^W=(\calf_t^W)_{t\geq 0}$.
We assume that $\mbb{F}^W$ is a subset of the full filtration; $\mbb{F}^W\subset \mbb{F}$.
\\
\\
For the ease of discussion, let us define the following spaces of the stochastic processes $(p\geq 1)$: \\
\bull $\mbb{S}_r^p(t,T)$ is the set of progressively measurable process $X$ taking values in $\mbb{R}^r$ and satisfying
\bea
\mbb{E}\Bigl[||X||_{[t,T]}^p\Bigr]:=\mbb{E}\Bigl[\sup_{s\in[t,T]}|X_s(\omega)|^p\Bigr]<\infty
\eea
where we use the notation
\be
||x||_{[a,b]}:=\sup\{|x_t|,t\in[a,b]\}
\ee
for $x:[0,T]\rightarrow \mbb{R}^r$. We write $||x||_{[0,t]}=||x||_{t}$. Its norm is defined by 
\be
||X||_{\mbb{S}^p_r(t,T)}:=\left\{\mbb{E}\Bigl[||X||_{[t,T]}^p
\Bigr]\right\}^{1/p}.
\ee
\bull $\mbb{H}_r^p(t,T)$ is the set of progressively measurable process $X$ taking values in $\mbb{R}^r$ and satisfying
\be
\mbb{E}\left[\left(\int_t^T |X_t|^2 dt\right)^{p/2}\right]<\infty,
\ee
and its norm is defined by
\bea
||X||_{\mbb{H}_r^p(t,T)}:=\left\{\mbb{E}\Bigl[\left(\int_t^T |X_s|^2 ds\right)^{p/2} \Bigr]\right\}^{1/p}~.
\eea
In every space, the subscript $r$ may be omitted if the associated dimension is clearly seen from the 
context.

\section{A single security case}
Firstly, let us summarize the standing assumptions.
They are obviously not the weakest ones but 
allow simple analysis and also do not make the model unrealistic in a practical setup.
Note that the definition of each variable
will appear along the discussions in the following sections.
\subsubsection*{$\mathbf{Assumption~A}$}
{\it
$\caln(\omega, dt,dz)$ is a random counting measure of a marked point process  
with a bounded support $K\subset\mbb{R}\backslash\{0\}$ for its mark $z$,
and $H$ is a counting process.
All the stochastic processes which do not jump by $\caln$ and $H$ are assumed to be $\mbb{F}^W$-adapted
and hence continuous. This $\mbb{F}^W$ adaptedness 
includes all the stochastic processes defined below.
\vspace{2mm}
\\
$(a_1)$ $S:\Omega\times[0,T]\rightarrow \mbb{R}$ is non-negative and $S\in\mbb{S}^4(0,T)$.\\
$(a_2)$ $b, l: \Omega\times[0,T] \rightarrow \mbb{R}$ and $b,l \in\mbb{S}^4(0,T)$. \\ 
$(a_3)$ $\Lambda(\cdot,\cdot): \Omega\times [0,T] \times \mbb{R} \rightarrow \mbb{R}$ is such that
$\Lambda(t,\cdot)(\omega)$ is a non-negative measurable function with bounded support $K\subset \mbb{R}\backslash\{0\}$ 
for every $t\in[0,T]$ and $\omega\in\Omega$, 
and that $\Lambda(\cdot,z)$ is a uniformly bounded $\mbb{F}^W$-adapted process for every $z\in K$. \\
$(a_4)$ $\wt{\gamma} :\Omega\times[0,T]\rightarrow \mbb{R}$ are uniformly bounded and non-negative.\\
$(a_5)$ $M,\wt{\eta}, \lambda: \Omega\times[0,T]\rightarrow \mbb{R}$ are uniformly bounded and strictly positive.\\
$(a_6)$ $\wt{\xi}:\Omega\rightarrow \mbb{R}$ is strictly positive, bounded and $\calf_T^W$-measurable. \\ 
$(a_7)$ $\beta: \Omega\times[0,T]\rightarrow \mbb{R}$ is uniformly bounded. \\
$(a_8)$ There is no simultaneous jump between $\caln$ and $H$.}
\\\\
$\bold{Notation:}$ {\it For a bounded variable $x$, we denote its upper bound by $\bar{x}$.}
\subsection{The market description}
\label{sec-market}
We are interested in a market maker, who has to face uncertain customer orders 
regarding the single specified security.
An extension to a portfolio management including multiple securities will be discussed 
in later sections.
As a {\it good} market maker, the firm accepts every customer order with a predefined
bid-offer spread. Although the spread can be dynamic depending on the external market variables such as
the security's volatility, it is supposed that the firm does not adjust it in order
to control the customer flows based on the firm's proprietary reasons.

The market maker is assumed to buy and sell the security through the two major trading venues.
The first venue is a standard exchange, where
the market maker carries out absolutely continuous trading.  
It is assumed, however, to incur a linear stochastic  price impact.
In addition, the participants of the exchange (partially) infer the inventory size of the 
market maker. They expect future buy/sell orders from it and adjust their positioning accordingly.
We assume that the aggregate effects of the participants change the market price 
by the amount proportional to the inventory size, which becomes another source of the
market maker's future trading costs.

The second venue is the aggregate of the OTC block trades with the firm's customers or 
the dark pools.  The market maker can buy/sell a block trade without directly affecting the market price, 
however, its timing is assumed to be uncertain.
Although we call this venue {\it the dark pool},
it actually means the aggregate of OTC block trades with the firm's counterparties 
as well as potentially multiple dark pools to which the market maker can access. It is a simplistic model 
for which we do not consider an order-size dependent intensity process nor possibility of the partial execution. 
Unfortunately, this seems unavoidable to keep the problem tractable.

In addition to the above two trading venues, 
the market maker can match an incoming customer order to its outstanding position being warehoused 
in its balance sheet. This is the distinguishing feature of the market maker.
Because this is the most profitable way to reduce the position, the market maker needs to adjust 
buy/sell orders based on the expected future customer flows.
If the market maker cannot answer an incoming customer order within its inventory, 
it needs to borrow the security through the corresponding repo market 
by paying the stochastic repo rate. On the other hand, when its inventory is positive, the market maker earns money 
by lending the security through the repo market.
\\

We model the the market-maker's position at time $s>t$ starting from the position size $x\in\mbb{R}$ at time $t$ as
\bea
X_s^{\pi,\del}(t,x)=x+\int_t^s \int_K z \caln(du,dz)+\int_t^s \pi_u du+\int_t^s \del_u dH_u
\eea
where the second term describes the customer flow, which is represented by the marked point process expressed by
the counting measure $\caln$.
$K\subset \mbb{R}\backslash\{0\}$ is a bounded support for the mark $z$ which gives the size and direction ($+$ or $-$)
of the order~\footnote{This boundedness can be interpreted as the maximum acceptable order size set by the market maker.}.
$(\pi,\del)$ denotes an $\mbb{F}$-predictable trading strategy of the market maker through the exchange and the dark pool,
respectively.
$H$ is the counting process, whose jump signals the happening of an execution event in the dark pool.
For simplicity, we assume no simultaneous jump between $\caln$ and $H$.
The negative position size $X^{\pi,\del}<0$ is always interpreted as a short position 
taken by the security borrowing through the repo market.
Let us assume the existence of the compensators $\Lambda$ 
for $\caln$ ($\lambda$ for $H$) so that
\bea
&&\int_0^t \int_K \wt{\caln}(ds,dz)=\int_0^t \int_K \Bigl(\caln(ds,dz)-\Lambda(s,z)dz ds\Bigr)\nn \\
&&\int_0^t d\wt{H}_s=\int_0^t \Bigl(dH_s-\lambda_s ds\Bigr)
\eea
for $t\in[0,T]$ are $\mbb{F}$-martingales. This also implies that an occurrence of a customer order and 
an execution in the dark pool are totally inaccessible.
For later convenience, let us define 
\bea
&&\Phi_t:=\int_K z \Lambda(t,z)dz, \quad \Psi_t:=\int_K |z|\Lambda(t,z)dz, \quad \Phi_{2,t}:=\int_K z^2 \Lambda(t,z)dz
\eea
for $t\in[0,T]$, which are the moments of the size of the customer orders. By Assumption A, the above 
processes are uniformly bounded.

The price observed in the exchange $\wt{S}^{\pi,\del}(t,x)$ i.e., 
the market price under the impact of the market maker's strategy
$(\pi,\del)$ starting from the position size $x$ at time $t$, is assumed to be given by
\bea
\wt{S}_s^{\pi,\del}(t,x)=S_s+M_s\pi_s-\beta_s X_s^{\pi,\del}(t,x)~
\label{eq-security}
\eea
for $s\in[t,T]$. The second term denotes the stochastic linear price impact, where $M$ is the $F^W$-adapted impact factor.
The last term denotes the aggregate impact from the market participants' reactions to the 
market maker's inventory size. 

Notice that we are not assuming the perfect observability of the
market maker's position $X$ to the other investors. It is likely that they can infer $X$ only vaguely.
Thus,  their reactions likely have a big noise, too. However, this noise part can easily be 
absorbed into the definition of $S$, {\it the unaffected price} of the security.
For the market maker's point of view, $X$ is directly observable and $\beta$ is simply 
its coefficient which can be obtained by the linear regression of the security price.
We model $\beta$ as a uniformly bounded  process
possibly  being correlated with other market variables, such as 
volatility of the security.
Due to the presence of the customer orders, the last term is not directly determined 
by the trading volume in the past and hence different from the standard 
model of the permanent price impact.
It is more closely related to the works on the {\it large trader's problem}
studied by Jarrow~(1992)~\cite{Jarrow}, Cvitani\'c \& Ma~(1996)~\cite{Cv-Ma},
and Bank \& Baum~(2004)~\cite{Bank}.
\\

We model the cash flow in the interval $]t,T]$ to the market maker with strategy $(\pi,\del)$ in the following way:
\bea
&&-\int_t^T \wt{S}^{\pi,\del}_s(t,x)\pi_s ds-\int_t^T \int_K \wt{S}^{\pi,\del}_{s-}(t,x)
\bigl(1-{\rm sgn}(z)b_s\bigr)z\caln(ds,dz)\nn\\
&&\quad-\int_t^T \Bigl(\bigl(S_{s}-\beta_sX_{s-}^{\pi,\del}(t,x)\bigr)\del_s+\wt{\eta}_s|\del_s|^2\Bigr)dH_s
+\int_t^T l_s X_s^{\pi,\del}(t,x)ds~.
\eea
Let us explain the economic meaning of each term below:\\
\bull (1st term) The cash flow from the trades through the exchange. \\
\bull (2nd term) The cash flow from accepting the customer orders with a (proportional) bid/offer spread $b$.\\
\bull (3rd term) The cash flow from the trades through the dark pool. \\
\bull (4th term) The cash flow from the security borrowing/lending with a repo rate $l$.\\

We need additional comments for the third term describing the trades with the dark pool.
Firstly, the basic transaction price is given by
\bea
S_{s}-\beta_sX_{s-}^{\pi,\del}(t,x)
\eea
which does not include the price impact from the continuous trading of the market maker.
Inclusion of $M_s\pi_s$ to the price could induce price manipulation, and more importantly,
the trading counterparties will not  accept expensive price caused by the 
market maker's {\it temporal} trading activity.
We also add the spread $\wt{\eta} |\del|$ to the above price~\footnote{Thus the additional cost is given by 
$\wt{\eta}|\del|^2$.} as a {\it premium} that
the market maker pays to the counterparty who has accepted a block trade.
\\

We consider $T\simleq 1$~(year) as a relevant time span for the control of the market maker.
More realistically, it can be a Quarter or a half year, and we neglect the net proceeds from 
a money market account for this time interval. This can be understood as a (nearly) zero interest rate, or equivalently, we can
interpret that the cost function (see below) is given in the discounted basis.
We are also interested in relatively liquid market in a sense that 
the borrowing and lending of the security is always possible as long as a given stochastic repo
rate is paid.
In a highly illiquid market, neither seamless execution of market orders in the
exchange nor a functioning repo market can be expected.

\subsection{The market maker's problem}
\begin{definition}
~\footnote{
It may not necessary to constrain the admissible strategies as Markovian with respect to $X$.
However, limiting the strategy space at this stage makes the following analysis much clearer.}
We define the admissible strategies $\calu$ by the set of $\mbb{F}$-predictable processes $(\pi,\del)$
that belong to $\mbb{H}^2(0,T)\times \mbb{H}^2(0,T)$ and also Markovian with respect to the position size,
i.e., they are expressed with some measurable functions $(f^\pi, f^\del)$ by
\bea
\pi_s=f^\pi(s,X_{s-}^{\pi,\del}(t,x)), \quad \del_s=f^{\del}(s,X_{s-}^{\pi,\del}(t,x))
\label{D-U-markov}
\eea
where, for $a\in\{\pi,\del\}$, $f^a: \Omega \times [0,T] \times \mbb{R}\rightarrow \mbb{R}$ 
and  $f^{a}(\cdot,x)$ is an $\mbb{F}^W$-adapted process for all $x\in\mbb{R}$. 
\end{definition}

We suppose that the market maker tries to solve the following optimization problem:
\bea
&&\wt{V}(t,x)={\rm ess}\inf_{(\pi,\del) \in \calu}
\mbb{E}\left[\LDis \right.
\wt{\xi} |X_T^{\pi,\del}(t,x)|^2+\int_t^T \wt{\gamma}_s |X_s^{\pi,\del}(t,x)|^2ds\nn \\
&&+\int_t^T \bigl(\wt{S}_s^{\pi,\del}(t,x)\pi_s-l_s X_s^{\pi,\del}(t,x)\bigr)ds
+\int_t^T \int_K \wt{S}_{s-}^{\pi,\del}(t,x)(1-{\rm sgn}(z)b_s)z\caln(ds,dz)\nn\\
&&+\int_t^T \Bigl(\bigl[S_s-\beta_s X_{s-}^{\pi,\del}(t,x)\bigr]\del_s+\wt{\eta}_s|\del_s|^2\Bigr)dH_s
\left.\LDis \Bigr|\calf_t \right]~.
\label{mm-problem-1}
\eea
The first two terms are introduced to give penalties for the outstanding position size.
It is natural to consider that $\wt{\xi}$ and $\wt{\gamma}$ are proportional to the variance 
of the price process of the security.
One may also want to take into account the regulatory costs arising from the outstanding position 
in the balance sheet.  It is possible by an appropriate modification of 
$\wt{\gamma}$ and $l$ as long as the relevant costs can be
reasonably approximated by a quadratic function with respect to the position 
size $X$. Note that the coefficients of the quadratic function can be stochastic.
\\

We can observe that the expectation in (\ref{mm-problem-1}) is finite for all $(\pi,\del)\in \calu$.
This can be easily checked by the fact $X^{\pi,\del}(t,x) \in \mbb{S}^2(t,T)$ and $\wt{S}^{\pi,\del}(t,x)\in \mbb{H}^2(t,T)$.
However, due to the 2nd order terms of $(\pi,\del)$ arising from $(-\beta X^{\pi,\del}\pi)$ and $(-\beta X^{\pi,\del}\del)$,
the cost function could be unbounded from below, and then the problem would be ill-defined.
In order to guarantee the well-posedness of the problem, we need additional assumptions.

Firstly, let us write the dynamics of the $\mbb{F}^W$-adapted bounded process $\beta$ as
\bea
d\beta_t=\mu^\beta_t dt+\sigma^\beta_t dW_t~.
\eea
Furthermore, we denote
\bea
\xi:=\wt{\xi}-\frac{\beta_T}{2}, \quad \eta:=\wt{\eta}+\frac{\beta}{2},\quad \gamma:=\wt{\gamma}+\frac{\mu^\beta}{2}~.
\label{shift-def}
\eea
\subsubsection*{$\bold{Assumption~B}$}
{\it
$(b_1)$ $\mu^\beta: \Omega\times[0,T]\rightarrow \mbb{R}, ~
\sigma^\beta: \Omega\times[0,T]\rightarrow \mbb{R}^d$ are uniformly bounded, and $\mbb{F}^W$-adapted. \\
$(b_2)$ $\gamma$ is non-negative $d\mbb{P}\otimes dt$-a.e..\\
$(b_3)$ There exists a constant $c>0$ such that $\xi\geq c$ a.s. and
$M,\eta~\geq c$ $d\mbb{P}\otimes dt$-a.e..
}

\begin{definition}
The cost function for the market maker with a given position size $x\in \mbb{R}$ at $t\in[0,T]$ is
\bea
&&J^{t,x}(\pi,\del)=\mbb{E}\left[\LDis \right.
\xi|X_T^{\pi,\del}(t,x)|^2+\int_t^T \Bigl(\gamma_s|X_s^{\pi,\del}(t,x)|^2+X_s^{\pi,\del}(t,x)(\beta_sb_s\Psi_s-l_s)\Bigr)ds\nn\\
&&\hspace{-7mm}+\int_t^T \Bigl( M_s\pi_s^2+\lambda_s \eta_s \del_s^2
+\bigl[(S_s+M_s\Theta_s)\pi_s+S_s\lambda_s\del_s\bigr]+\bigl(S_s\Theta_s+\frac{\beta_s}{2}\Phi_{2,s}\bigr)
\Bigr)ds \left.\LDis \Bigr|\calf_t^W \right]
\eea
where $\Theta:\Omega\times[0,T]\rightarrow \mbb{R}$
is defined by $\Theta_s:=\Phi_s-b_s\Psi_s$.
\end{definition}

\begin{proposition} Under Assumptions A and B, the market maker's problem (\ref{mm-problem-1}) 
is equivalent to
\bea
&&V(t,x)={\rm ess}\inf_{(\pi,\del)\in\calu}J^{t,x}(\pi,\del)
\label{mm-problem-2}
\eea
and it has a unique optimal solution $(\pi^*,\del^*)\in\calu$.
\label{P-mm-problem}
\begin{proof}
Applying It\^o-formula, one obtains
\bea
&&-\int_t^T \beta_s X_{s-}^{\pi,\del}(t,x)dX_s^{\pi,\del}(t,x)=-\frac{\beta_T}{2}|X_T^{\pi,\del}(t,x)|^2
+\frac{\beta_t}{2}x^2+\int_t^T\frac{\beta_s}{2}\del_s^2 dH_s\nn \\
&&\qquad +\int_t^T\int_K \frac{\beta_s}{2}z^2\caln(ds,dz)
+\int_t^T \frac{1}{2}|X_s^{\pi,\del}(t,x)|^2(\mu^\beta_s ds+\sigma^\beta_s dW_s)~.\nn
\eea
Then, replacing the $\beta$-proportional terms in (\ref{mm-problem-1}) $\Bigl(-\int_t^T \beta_s X_{s-}^{\pi,\del}(t,x)[\pi_s ds+\del_s dH_s]\Bigr)$ by using the above relation and (\ref{shift-def}) yields
\bea
&&\wt{V}(t,x)-\frac{\beta_t}{2}x^2={\rm ess}\inf_{(\pi,\del)\in \calu}
\mbb{E}\left[\LDis \right. \xi|X_T^{\pi,\del}(t,x)|^2+\int_t^T \Bigl(
\gamma_s|X_s^{\pi,\del}(t,x)|^2+X_s^{\pi,\del}(t,x)(\beta_s b_s\Psi_s-l_s)\Bigr)ds\nn \\
&&\quad+\int_t^T\left\{M_s \pi_s^2+\lambda_s \eta_s \del_s^2+
\bigl[(S_s+M_s\Theta_s)\pi_s+S_s\lambda_s\del_s\bigr]+\bigl(S_s\Theta_s+\frac{\beta_s}{2}\Phi_{2,s}\bigr)\right\}ds
\left.\LDis \Bigr|\calf_t\right]
\label{eq-vtil-shift}
\eea
where the integrals by the counting measures are replaced by their compensators.
Here, one can check that the local martingales are true martingales under the assumptions.
In particular, one can use the Burkholder-Davis-Gundy (BDG) inequality,
the fact that $X^{\pi,\del}\in\mbb{S}^2(t,T)$ and the boundedness of $\sigma_\beta$ for the $dW$ integration term.
For the jump part, it suffices to check that the integration by the corresponding compensator is 
in $\mbb{L}^1(\Omega)$ (See, for example, Corollary C4, Chapter VIII in \cite{Bremaud}.), 
which can be confirmed by the boundedness of the 
compensators, $\lambda$ and $\Lambda$.

Because all the processes except $(X^{\pi,\del},\pi,\del)$ are $\mbb{F}^W$-adapted
and $(\pi,\del)\in \calu$ satisfies (\ref{D-U-markov}),
the expectation conditioned on $\calf_t$ in (\ref{eq-vtil-shift}) can be replaced by $\calf_t^W\vee\sigma\{X_t^{\pi,\del}\}$
thanks to the Markovian nature of $X^{\pi,\del}$. Notice that the information of the counting measures $\caln, H$ only 
appears through the position size $X^{\pi,\del}$. However, $X_t^{\pi,\del}(t,x)=x$ has been already fixed.
Thus, redefining the value function by $\displaystyle V(t,x):=\wt{V}(t,x)-\frac{\beta_t}{2}x^2$, one obtains the result
(\ref{mm-problem-2}) as an equivalent problem for the maker maker.

The remaining claims easily follow from the standard arguments (See, for example, Theorem 3.1 in Bismut~(1976)~\cite{Bismut}.)
since now all the quadratic terms have positive coefficients. 
For simplicity, let us consider the case where the initial time is zero, $t=0$. 
The cost function $J^{0,x}$ is a continuous map from $\calu$ to $\mbb{R}$ and obviously strictly convex.
It is also proper since, for example,  $J^{0,x}(0)<\infty$. 
We also have the so-called coerciveness since
\be
J^{0,x}(u)\nearrow \infty, \quad {\rm when}\quad ||u||_{\mbb{H}^2_2(0,T)} \nearrow \infty~.
\ee
The above observations and the fact that $\calu$ is a Hilbert space tell us that,
for  a large enough $\alpha\in\mbb{R}$, the set $\{u\in\calu: J^{0,x}(u)\leq \alpha\}$
is non-empty, convex and weakly-compact. Thus, there exists an minimizer, which is unique due to 
the strict convexity of the cost function.
\end{proof}
\end{proposition}

\subsubsection*{Remarks on $\beta X^{\pi,\del}$ in (\ref{eq-security}) }
The presence of $\beta X^{\pi,\del}$ term in (\ref{eq-security}) is not necessarily appropriate for every type of investors.
For example, suppose that the investor is risk-neutral and $\beta$ is positive.
In this case, the investor may accumulate an extremely large long position which 
would make the security price significantly negative. The investor can receive 
positive cash flow by further increasing her long position which makes the system ill-defined.
However, as we have seen in the above discussion, it does not cause any 
regularity problem under mild conditions regarding the penalty size on the 
outstanding position of the investor.
Although one may feel uneasy by the fact that the well-posedness of the model depends on the 
risk-averseness of the agent, we think that this term makes the model more realistic for the market maker.
In fact, this term is expected to arise exactly because the other investors know 
that the relevant market maker has to operate with a rather stringent position limit i.e., risk averse.
From the view point of the market maker, it is being squeezed
by the other investors as long as there exists an information leak about its position size.

\section{Solving the problem}
\subsection{A candidate solution}
Let us prepare the optimality principle for the current problem.
\begin{proposition} (Optimality Principle) Let Assumptions A and B are satisfied. Then,  \\
(a) For all $x\in\mbb{R}$, $(\pi,\del)\in\calu$ and $t\in[0,T]$, the process 
\bea
&&\left(\LDis \right. V(s,X_s^{\pi,\del}(t,x))+\int_t^s
\Bigl(\gamma_u |X_u^{\pi,\del}(t,x)|^2+X_u^{\pi,\del}(t,x)\bigl(\beta_u b_u\Psi_u-l_u\bigr)\Bigr)du\nn \\
&&\quad+\int_t^s \Bigl( M_u\pi_u^2+\lambda_u \eta_u \del_u^2
+\bigl[(S_u+M_u\Theta_u)\pi_u+S_u\lambda_u\del_u\bigr]+\bigl(S_u\Theta_u+\frac{\beta_u}{2}\Phi_{2,u}\bigr)
\Bigr)du \left.\LDis \right)_{s\in[t,T]}\nn
\eea
is an $\mbb{F}$-submartingale. \\
(b) $(\pi^*,\del^*)$ is optimal if and only if 
\bea
&&\left(\LDis \right. V(s,X_s^{\opi,\odel}(t,x))+\int_t^s
\Bigl(\gamma_u |X_u^{\opi,\odel}(t,x)|^2+X_u^{\opi,\odel}(t,x)\bigl(\beta_u b_u\Psi_u-l_u\bigr)\Bigr)du\nn \\
&&\quad+\int_t^s \Bigl( M_u{\opi}_u^2+\lambda_u \eta_u {\odel}_u^2
+\bigl[(S_u+M_u\Theta_u)\opi_u+S_u\lambda_u\odel_u\bigr]+\bigl(S_u\Theta_u+\frac{\beta_u}{2}\Phi_{2,u}\bigr)
\Bigr)du \left.\LDis \right)_{s\in[t,T]}\nn
\eea
is an $\mbb{F}$-martingale.
\label{P-optimality}
\begin{proof}
One can easily confirm it from the definition of the value function $V$, 
the fact that $V(T,X_T^{\pi,\del})=\xi|X_T^{\pi,\del}|^2$
and the form of the cost function $J^{t,x}(\pi,\del)$.  See, for example, Proposition (A.1) of 
Mania \& Tevzadze (2003)~\cite{Mania}.
\end{proof}
\end{proposition}

Firstly, by following the method proposed by Mania~\& Tevzadze~\cite{Mania}, 
we derive the BSDEs from the necessary condition so that the above optimality principle is satisfied.
Then, we are going to show that there exists a solution for every BSDE and confirm 
that it actually satisfies the optimality principle.
This gives us one optimal solution. But we know that the solution is also unique due to Proposition~\ref{P-mm-problem}.

Let us assume that the $\mbb{F}^W$ semimartingale $\Bigl(V(t,x)\Bigr)_{t\in[0,T]}$ 
has the following decomposition for every $x\in \mbb{R}$:
\bea
\label{eq-st-HJB}
V(s,x)=V(t,x)+\int_t^s a(u,x)du+\int_t^sZ(u,x)dW_u
\eea
where $a:\Omega\times[0,T]\times \mbb{R}\rightarrow \mbb{R},~~Z:\Omega\times [0,T]\times \mbb{R}\rightarrow \mbb{R}^d$
and $a(\cdot,x)$ as well as $Z(\cdot,x)$ are $\mbb{F}^W$-adapted processes for all $x\in\mbb{R}$.
Let us suppose $V(t,x)$ are twice differentiable with respect to $x$.
By applying It\^o-Ventzell formula, we obtain
\bea
&&V(s,X_s^{\pi,\del}(t,x))=V(t,x)+\int_t^s a(u,X_u^{\pi,\del}(t,x))du+\int_t^s Z(u,X_u^{\pi,\del}(t,x))dW_u\nn \\
&&+\int_t^s V_x(u,X_u^{\pi,\del}(t,x))\pi_u du+
\int_t^s \int_K \Bigl(V(u,X_{u-}^{\pi,\del}(t,x)+z)-V(u,X_{u-}^{\pi,\del}(t,x))\Bigr)\caln(du,dz)\nn \\
&&+\int_t^s \Bigl(V(u,X_{u-}^{\pi,\del}(t,x)+\del_u)-V(u,X_{u-}^{\pi,\del}(t,x))\Bigr)dH_u~.
\eea
Separating the local martingale parts, a necessary condition for the 
optimality principle is given by
\bea
&&a(u,x)+\int_K\Bigl(V(u,x+z)-V(u,x)\Bigr)\Lambda(u,z)dz+\gamma_u x^2+x(\beta_u b_u \Psi_u-l_u)\nn \\
&&+\Bigl(S_u\Theta_u+\frac{\beta_u}{2}\Phi_{2,u}\Bigr)+\inf_{\pi,\del}\left\{\LDis \right.
V_x(u,x)\pi+\bigl(V(u,x+\del)-V(u,x)\bigr)\lambda_u\nn \\
&&+M_u\pi^2+\lambda_u \eta_u \del^2 +(S_u+M_u\Theta_u)\pi+S_u\lambda_u \del \left.\LDis \right\}
=0
\label{a-condition}
\eea 
$d\mbb{P}\otimes dt$-a.e. in $\Omega\times[0,T]$ for every $x\in\mbb{R}$.

Substituting the resultant drift term $a(\cdot,\cdot)$ into (\ref{eq-st-HJB})
yields a backward stochastic PDE
\bea
&&V(t,x)=\xi|x|^2+\int_t^T\left\{ 
\int_K \bigl[V(u,x+z)-V(u,x)\bigr]\Lambda(u,z)dz+\gamma_u x^2+x(\beta_u b_u\Psi_u-l_u)\right. \nn \\
&& \qquad\quad \left.+\Bigl(S_u\Theta_u+\frac{\beta_u}{2}\Psi_{2,u}\Bigr)\right\}du
+\int_t^T \inf_{\pi,\del}\left\{\Bigl. V_x(u,x)\pi+\bigl[V(u,x+\del)-V(u,x)\bigr]\lambda_u \right.\nn \\
&&\qquad\quad \left.\Bigl.+M_u\pi^2+\lambda_u\eta_u \del^2+(S_u+M_u\Theta_u)\pi+S_u\lambda_u\del\right\}du
-\int_t^T Z(u,x)dW_u,
\label{eq-bspde}
\eea
which is sometimes called a stochastic HJB equation. $(\pi,\del)$ should be chosen for each $u\in[t,T]$.
Exploiting the quadratic nature, let us hypothesize that, for every $t\in[0,T]$ and $x\in \mbb{R}$,
\bea
&&V(t,x)=V_2(t)x^2+2V_1(t)x+V_0(t) \\
&&Z(t,x)=Z_2(t)x^2+2Z_1(t)x+Z_0(t)
\eea
where $V_2, V_1, V_0: \Omega\times[0,T]\rightarrow \mbb{R}$ and
$Z_2,Z_1,Z_0: \Omega\times[0,T]\rightarrow \mbb{R}^d$ are $\mbb{F}^W$-adapted processes.
Then, (\ref{a-condition}) can be rewritten as
\bea
&&a(u,x)+\bigl(2V_2(u)\Phi_u x+V_2(u)\Phi_{2,u}+2V_1(u)\Phi_u\bigr)+\gamma_u x^2+x\bigl(\beta_u b_u \Psi_u-l_u\bigr)+\Bigl(S_u\Theta_u+\frac{\beta_u}{2}\Phi_{2,u}\Bigr)\nn \\
&&+\inf_{\pi,\del}\left\{\LDis \right.
M_u\Bigl(\pi+\frac{\bigl[V_2(u)x+V_1(u)+\frac{1}{2}(S_u+M_u\Theta_u)\bigr]}{M_u}\Bigr)^2\nn \\
&&\quad +\lambda_u\bigl[V_2(u)+\eta_u\bigr]\Bigl(\del+\frac{\bigl[V_2(u)x+V_1(u)+\frac{1}{2}S_u\bigr]}{V_2(u)+\eta_u}
\Bigr)^2\nn  \\
&&\quad-\frac{1}{M_u}\Bigl(V_2(u)x+V_1(u)+\frac{1}{2}(S_u+M_u\Theta_u)\Bigr)^2-\lambda_u\frac{\bigl[V_2(u)x+V_1(u)+\frac{1}{2}S_u\bigr]^2}{V_2(u)+\eta_u} \left.\LDis\right\}\nn \\
&&=0~~d\mbb{P}\otimes dt-a.e. .
\eea
For the well-posedness, we must have $V_2+\eta>0$~$d\mbb{P}\otimes dt$-a.e..

Gathering each of $(x^2,x^1,x^0)$-proportional terms in (\ref{eq-bspde}), one obtains the following result.
\subsubsection*{A Candidate Solution}
{\it
A ``candidate" of the optimal solution and the corresponding value function 
for the market maker's problem (\ref{mm-problem-2})
are given by
\bea
\label{eq-opi}
&&\pi_u^*=-\frac{1}{M_u}\Bigl(V_2(u)X_{u-}^{\opi,\odel}(t,x)+V_1(u)+\frac{1}{2}(S_u+M_u\Theta_u)\Bigr)  \\
\label{eq-odel}
&&\del_u^*=-\frac{\displaystyle \bigl[V_2(u)X_{u-}^{\opi,\odel}(t,x)+V_1(u)+\frac{1}{2}S_u\bigr]}{V_2(u)+\eta_u} 
\eea
for $u\in[t,T]$ and $V(t,x)=V_2(t)x^2+2V_1(t)x+V_0(t)$, respectively. Here,  $X^{\opi,\odel}(t,x)$
is the solution of 
\be
X_s^{\opi,\odel}(t,x)=x+\int_t^s \int_K z\caln(du,dz)+\int_t^s \pi_u^* du+\int_t^s \del_u^* dH_u,~s\in[t,T]~.
\label{eq-xopt}
\ee
$(V_2,Z_2), (V_1,Z_1)$ and $(V_0,Z_0)$ 
must be the well-defined solutions of
the following three BSDEs
\bea
\label{eq-v2}
&&V_2(t)=\xi+\int_t^T \left\{-\Bigl(\frac{1}{M_u}+\frac{\lambda_u}{V_2(u)+\eta_u}\Bigr)V_2(u)^2+\gamma_u\right\}du
-\int_t^T Z_2(u)dW_u 
\eea
\vspace{-3mm}
\bea
\label{eq-v1}
&&V_1(t)=-\int_t^T \left\{\LDis \right. V_2(u)\Bigl(\frac{1}{M_u}+\frac{\lambda_u}{V_2(u)+\eta_u}\Bigr)V_1(u)
-\frac{1}{2}\bigl(\beta_u b_u \Psi_u-l_u\bigr)\nn \\
&&\quad+V_2(u)\left( \Bigl[\frac{1}{M_u}+\frac{\lambda_u}{V_2(u)+\eta_u}\Bigr]\frac{S_u}{2}
-\frac{1}{2}\Theta_u-b_u\Psi_u\right)\left.\LDis \right\}du-\int_t^T Z_1(u)dW_u 
\eea
\vspace{-3mm}
\bea
\label{eq-v0}
&&V_0(t)=-\int_t^T \left\{\LDis \right. \Bigl(\frac{1}{M_u}+\frac{\lambda_u}{V_2(u)+\eta_u}\Bigr)
\Bigl(V_1(u)+\frac{S_u}{2}\Bigr)^2-V_1(u)\bigl(\Phi_u+b_u\Psi_u\bigr)\nn \\
&&\qquad-V_2(u)\Phi_{2,u}-\frac{1}{2}(S_u\Theta_u+\beta_u\Phi_{2,u})+\frac{1}{4}M_u\Theta_u^2
\left.\LDis\right\}du-\int_t^T Z_0(u)dW_u~,
\eea
satisfying 
\be
V_2+\eta>0
\ee
$d\mbb{P}\otimes dt$-a.e. in $\Omega\times[0,T]$.
}

\subsection{Verification}
We are now going to study each BSDE and show the existence of the candidate solution,
and also confirm that it actually satisfies  the optimality principle.

\begin{proposition}
Under Assumptions A and B, the BSDE (\ref{eq-v2}) has a unique solution in $(V_2,Z_2)\in \mbb{S}^p(0,T)\times \mbb{H}^p_d(0,T)$
for $\forall~p>1$, and in particular $V_2(t)$ satisfies for every $t\in[0,T]$ and $\ep>0$ that:
\bea
&&\frac{1}{\displaystyle \mbb{E}\left[\frac{1}{\xi}+\int_t^T\Bigl(\frac{1}{M_s}+\frac{\lambda_s}{\eta_s}\Bigr)ds
\Bigr|\calf^W_t\right]}\leq V_2(t) \nn \\
&&\hspace{15mm} \leq  \frac{1}{(T-t+\ep)^2}\mbb{E}\left[\ep^2\xi +
\int_t^T \Bigl(M_s+(T-s+\ep)^2\gamma_s\Bigr)ds\Bigr|\calf_t^W\right]~.
\eea
\label{P-v2-bound}
\begin{proof}
Let us define the function as
\be
f(t,y)=-\Bigl(\frac{1}{M_t}+\frac{\lambda_t}{y+\eta_t}\Bigr)y^2+\gamma_t~.
\ee
Firstly, let us consider the BSDE
\bea
Y_t=\xi+\int_t^T f(s,Y_s\vee 0)ds-\int_t^T Z_s dW_s 
\label{eq-y0}
\eea
Due to Assumptions A, B and the definitions (\ref{shift-def}), $\xi$ and $f(t,y\vee 0)$ with any fixed $y$ 
are bounded. Furthermore, it is clear that $f(t,y\vee 0)$ is a decreasing function in $y$.
Thus, (\ref{eq-y0}) satisfies the standard monotone conditions for the BSDE.
By Theorem 5.27 in~\cite{Pardoux}~\footnote{One can simply put $\mu$=$l$=0 in the theorem.}, 
there exists a unique solution 
$(Y,Z)\in \mbb{S}^p(0,T)\times \mbb{H}_d^p(0,T)$ for all $p>1$.
On the other hand, it is clear that we have a trivial solution $(Y,Z)=(0,0)$ if $\xi=0$ and $\gamma=0$.
Since the terminal value and the driver $f$ is increasing in $(\xi,\gamma)$,
we actually have $Y\geq 0$ by the comparison theorem (See, for example, Proposition 5.33 in \cite{Pardoux}.). 
As a result, the BSDE (\ref{eq-v2}) has
a unique solution $(V_2,Z_2)\in \mbb{S}^{p}(0,T)\times \mbb{H}^p_d(0,T)$ for all $p>1$,
and in addition, $V_2$ is non-negative.

The derivation of the upper and lower bounds is an adaptation of Proposition 2.1 in
Ankirchner, Jeanblanc \& Kruse~(2014)~\cite{Jeanblanc} for our problem. 
Let us start from the derivation of the upper bound.
For all $y,k \in \mbb{R}$, 
\be
y^2-2ky+k^2\geq 0
\ee
is satisfied. For an arbitrary constant $\ep>0$, choosing $\displaystyle k=\frac{M_t}{T-t+\ep}$ 
yields
\bea
-\Bigl(\frac{1}{M_t}+\frac{\lambda_t}{y+\eta_t}\Bigr)y^2\leq -\frac{1}{M_t}y^2
\leq -\frac{2}{T-t+\ep}y+\frac{M_t}{(T-t+\ep)^2}
\label{driver-ineq}
\eea
for all $y\geq 0$.

With some abuse of notation, consider the next linear BSDE 
\bea
Y_t^\ep=\xi+\int_t^T\left\{-\frac{2}{T-s+\ep}Y_s^\ep+\frac{M_s}{(T-s+\ep)^2}+\gamma_s\right\}ds-\int_t^T Z_s^\ep dW_s~.
\eea
This is a linear BSDE with a bounded Lipschitz constant. Due to the boundedness of $\xi,M,\gamma$,
there exists a unique solution $(Y^\ep,Z^\ep)\in \mbb{S}^p(0,T)\times \mbb{H}^p_d(0,T)$ for all $p>1$.
By the inequality (\ref{driver-ineq}) and the comparison theorem, we have
\bea
V_2(t)\leq Y_t^\ep
\eea
for all $t\in[0,T]$ and $\ep>0$. In addition, $Y^\ep$ can be solved as
\bea
&&Y_t^\ep=\mbb{E}\left[\xi e^{-\int_t^T \frac{2}{T-s+\ep}ds}+
\int_t^T e^{-\int_t^s \frac{2}{T-u+\ep}du}\Bigl(
\frac{M_s}{(T-s+\ep)^2}+\gamma_s\Bigr)ds\Bigr|\calf_t^W\right]\nn \\
&&=\frac{1}{(T-t+\ep)^2}\mbb{E}\left[\ep^2\xi
+\int_t^T \Bigl(M_s+(T-s+\ep)^2\gamma_s\Bigr)ds\Bigr|\calf_t^W\right]
\eea
and hence we obtained the desired upper bound.

Now, let us study the lower bound.
Put
\be
\wt{V}_t:=\mbb{E}\left[\frac{1}{\xi}+\int_t^T \Bigl(\frac{1}{M_s}+\frac{\lambda_s}{\eta_s}\Bigr)ds\Bigr|\calf_t^W\right].
\ee
Due to the existence of a constant $c>0$ such that $\xi,M,\eta\geq c$, 
it satisfies
\bea
\frac{1}{\bigr.\bar{\xi}} \leq \wt{V} \leq \frac{1}{c}\Bigl(1+T(1+\bar{\lambda})\Bigr)~(:=\kappa)
\eea
where $\bar{\xi},\bar{\lambda}$ are the upper bounds of $\xi,\lambda$, respectively.
Therefore, there exists $\wt{Z}\in \mbb{H}^p_d(0,T),~\forall p>0$ such that
\bea
d\wt{V}_t=-\Bigl(\frac{1}{M_t}+\frac{\lambda_t}{\eta_t}\Bigr)dt+\wt{Z}_tdW_t~.
\eea
Then the process $\wt{U}_t:=1/\wt{V}_t$, which has the bounds $1/\kappa \leq \wt{U}\leq \bar{\xi}$,
satisfies
\bea
\wt{U}_t=\xi+\int_t^T\left\{-\Bigl(\frac{1}{M_s}+\frac{\lambda_s}{\eta_s}\Bigr)\wt{U}_s^2
-\frac{|\Gamma_s|^2}{\wt{U}_s}\right\}ds-\int_t^T \Gamma_s dW_s~,
\eea
where $\displaystyle \Gamma:=-\frac{\wt{Z}}{\wt{V}^2}$.
Here, the terminal value, the first term of the driver  and the coefficient of $|\Gamma|^2$
are all bounded. Thus the comparison theorem for the quadratic BSDE 
(See, Theorem 2.6 in \cite{Kobylanski}), one sees $\wt{U}_t\leq V_2(t)$ and 
hence the desired result is obtained.
\end{proof}
\end{proposition}

Since the BSDEs for $(V_1,Z_1)$ and $(V_0,Z_0)$ are linear, one can use
popular established results to obtain the  next Proposition.
\begin{proposition}
Under Assumptions A and B, there exist unique solutions $(V_1,Z_1)\in \mbb{S}^4(0,T)\times \mbb{H}^4_d(0,T)$ for 
(\ref{eq-v1}), and $(V_0,Z_0)\in \mbb{S}^2(0,T)\times \mbb{H}_d^2(0,T)$ for (\ref{eq-v0}),
respectively.
\label{P-v1-v0}
\begin{proof}
We denote by $C$ some positive constant, which may change line by line.
From Proposition~\ref{P-v2-bound}, $V_2$ is uniformly bounded and hence so is the linear coefficient of $V_1$.
In addition,
\bea
&&\mbb{E}\left[\LDis \left( \int_0^T \left|
V_2(s)\Bigl(\Bigl[\frac{1}{M_s}+\frac{\lambda_s}{V_2(s)+\eta_s}\Bigr]\frac{S_s}{2}-
\frac{\Theta_s}{2}-b_s\Psi_s\Bigr)-\frac{1}{2}(\beta_s b_s \Psi_s-l_s)\right|ds\right)^4\right]\nn \\
&&\leq C\mbb{E}\Bigl[
1+\int_0^T\Bigl(|S_s|^4+|b_s|^4+|l_s|^4\Bigr)ds\Bigr]<\infty
\eea
Thus, by Theorem 5.21 (see also Section 5.3.5) in \cite{Pardoux}, there exists a 
unique solution $(V_1,Z_1)\in\mbb{S}^4(0,T)\times \mbb{H}^4_d(0,T)$.

As for (\ref{eq-v0}), it is easy to see that
\bea
&&\mbb{E}\left[\LDis \left(\int_0^T\left|
\Bigl(\frac{1}{M_s}+\frac{\lambda_s}{V_2(s)+\eta_s}\Bigr)\Bigl(V_1(s)+\frac{S_s}{2}\Bigr)^2
-V_1(s)(\Phi_s+b_s\Psi_s)-V_2(s)\Phi_{2,s} \right.\right. \right. \nn \\
&&\quad\left.\LDis \left.\left.-\frac{1}{2}(S_s\Theta_s+\beta_s\Phi_{2,s})+\frac{1}{4}M_s\Theta_s^2\right|ds\right)^2\right]\nn \\
&&\leq C\mbb{E}\left[1+\int_0^T\Bigl(|V_1(s)|^4+|S_s|^4+|b_s|^4\Bigr)ds\right]<\infty,
\eea
where we have used $V_1\in\mbb{S}^4(0,T)$ proved in the previous arguments.
Thus, by the same reasoning, there exists a unique solution $(V_0,Z_0)\in\mbb{S}^2(0,T)\times \mbb{H}^2_d(0,T)$.
\end{proof}
\end{proposition}

In order to check the optimality condition, we also need the following property of $X^{\opi,\odel}$.
\begin{proposition}
Under Assumptions A and B, the process of the position size $\Bigl(X^{\opi,\odel}_s(t,x)\Bigr)_{s\in[t,T]}$ given by (\ref{eq-xopt}) belongs to $\mbb{S}^4(t,T)$.
\label{P-x}
\begin{proof}
Let us take the starting time $0$  and write $X_s^{\opi,\odel}(0,x)$ as $X_s^*$ for simplicity. Then,
\bea
&&X_s^{\opi,\odel}(0,x)=x+\int_0^s\int_K z\wt{\caln}(du,dz)-\int_0^s\frac{
\bigl(V_2(u)X_{u-}^{\opi,\odel}(0,x)+V_1(u)+\frac{S_u}{2}\bigr)}{V_2(u)+\eta_u}d\wt{H}_u \nn \\
&&\quad-\int_0^s\left\{\LDis \right. V_2(u)\Bigl(\frac{1}{M_u}+\frac{\lambda_u}{V_2(u)+\eta_u}\Bigr)X_u^{\opi,\odel}(0,x)
+\Bigl(\frac{1}{M_u}+\frac{\lambda_u}{V_2(u)+\eta_u}\Bigr)\Bigl(V_1(u)+\frac{S_u}{2}\Bigr)\nn \\
&&\qquad+\frac{\Theta_u}{2}-\Phi_u\left.\LDis \right\}du
\eea
Under Assumptions A and B, there exists some positive constant $C$ such that
\bea
&&|X_t^*|^4 \leq 
C\left[1+ \int_0^t\Bigl(|X_u^*|^4+|V_1(u)|^4+|S_u|^4+|b_u|^4 \Bigr)du \right. \nn  \\
&&\quad\left.+\Bigl(\int_0^t\int_K z\wt{\caln}(du,dz)\Bigr)^4+
\left(\int_0^t \frac{[V_2(u)X_{u-}^*+V_1(u)+\frac{S_u}{2}]}{V_2(u)+\eta_u}d\wt{H}_u\right)^4
\right]~,
\label{eq-x4-org}
\eea
for every $t\in[0,T]$.
Let us define a sequence of $\mbb{F}$-stopping times $(\tau_n)_{n\geq 0}$ by
\bea
\tau_n:=\inf\left\{t\geq 0: |X_t^*| > n\right\}\wedge T~,
\eea
and denote the $\tau_n$-stopped process of the position size as $X^{*\tau_n}_s=X^*_{s\wedge \tau_n}$.
Since we already know that $X^{*}\in\mbb{S}^2(0,T)$, it is clear that $\tau_n \rightarrow T$ a.s. as $n\rightarrow \infty$.

The BDG inequality (see, for example, Theorem 10.36 in \cite{hwy} for general local martingales) 
and positivity of integrand yield
\bea
&&\mbb{E}\Bigl[|X_t^{*\tau_n}|^4\Bigr]\leq C\mbb{E}\left[
1+\int_0^t \Bigl(|X_u^{*\tau_n}|^4+|V_1(u)|^4+|S_u|^4+|b_u|^4\Bigr)du \right. \nn \\
&&\qquad+\left(\int_0^t\int_K z^2\caln(du,dz)\right)^2 \left.+\left(\int_0^t \Bigl|\frac{V_2(u)X_{u-}^{*\tau_n}+V_1(u)+\frac{S_u}{2}}{V_2(u)+\eta_u}\Bigr|^2dH_u\right)^2
\right]. 
\label{eq-x4th}
\eea
Again by the BDG inequality and the boundedness of $\lambda$, one has
\bea
\mbb{E}\left[\Bigl(\int_0^t |X_{u-}^{*\tau_n}|^2dH_u\Bigr)^2\right]
&&\leq 2\mbb{E}\left[\Bigl(\int_0^t |X_{u-}^{*\tau_n}|^2d\wt{H}_u\Bigr)^2\right]
+2\mbb{E}\left[\Bigl(\int_0^t |X_{u-}^{*\tau_n}|^2\lambda_u du\Bigr)^2\right] \nn \\
&& \leq C\mbb{E}\left[\int_0^t |X_{u}^{*\tau_n}|^4 du\right]~.
\eea
One obtains, by similar analysis, that
\bea
\mbb{E}\left[
\left(\int_0^t \Bigl|\frac{V_2(u)X_{u-}^{*\tau_n}+V_1(u)+\frac{S_u}{2}}{V_2(u)+\eta_u}\Bigr|^2dH_u\right)^2
\right]\leq
C\mbb{E}\left[ \int_0^t \Bigl(|V_1(u)|^4+|S_u|^4\Bigr)du+\int_0^t |X_u^{*\tau_n}|^4 du\right]~.
\eea
Thus, it can be shown from (\ref{eq-x4th}) and the boundedness of $K$ that 
\be
\mbb{E}\Bigl[|X_t^{*\tau_n}|^4\Bigr]\leq C\mbb{E}\left[
1+||V_1||^4_T+||S||^4_T+||b||^4_T+\int_0^t |X_u^{*\tau_n}|^4 du \right]
\ee
and hence, by the Gronwall lemma, for $\forall t\in[0,T]$,
\bea
\mbb{E}\Bigl[|X_t^{*\tau_n}|^4\Bigr]\leq C\mbb{E}\left[
1+||V_1||^4_T+||S||^4_T+||b||^4_T\right]e^{CT}<\infty~.
\eea
Passing to the limit $n\rightarrow \infty$, we see $\mbb{E}\Bigl[|X_t^{*}|^4\Bigr]<C$ 
for every $t\in[0,T]$ with some positive constant $C$.
Using the BDG inequality and the above estimate, we obtain from (\ref{eq-x4-org}) that
\bea
\mbb{E}\Bigl[||X^*||^4_T\Bigr]
\leq C\mbb{E}\left[1+||V_1||^4_T+||S||^4_T+||b||^4_T+\int_0^T |X_u^*|^4 du\right]<\infty~.
\eea
\end{proof}
\end{proposition}

\begin{corollary}
Under Assumptions A and B, the candidate solution $(\opi,\odel)$ given by (\ref{eq-opi}) and (\ref{eq-odel})
is well-defined, unique and satisfies $(\opi,\odel)\in \mbb{S}^4(t,T)\times \mbb{S}^4(t,T)\subset \calu$.
\label{C-opt}
\end{corollary}

Finally, we arrived the first main result of the paper.
\begin{theorem}
Under Assumptions A and B, the candidate solution $(\opi,\odel)$ given by (\ref{eq-opi}) and (\ref{eq-odel})
is, in fact, the unique optimal solution of the market maker's problem given by (\ref{mm-problem-2}).
\label{T-main}
\begin{proof}
It suffices to confirm that the optimality principle of Proposition~\ref{P-optimality} is indeed satisfied.
Firstly, we have to see
\bea
&&\bullet~ \left(\int_t^s Z(u,X_u^{\opi,\odel}(t,x))dW_u\right)_{s\in[t,T]}\nn \\
&&\bullet~ \left(\int_t^s \Bigl(V(u,X_{u-}^{\opi,\odel}(t,x)+\del_u^*)-V(u,X_{u-}^{\opi,\odel}(t,x))\Bigr)d\wt{H}_u 
\right)_{s\in[t,T]}\nn \\
&&\bullet~ \left(\int_t^s \int_K \Bigl(V(u,X_{u-}^{\opi,\odel}(t,x)+z)-V(u,X_{u-}^{\opi,\odel}(t,x))\Bigr)\wt{\caln}(du,dz)
\right)_{s\in[t,T]}\nn
\eea
are all true $\mbb{F}$-martingales.
For notational simplicity, let us put $t=0$ and $X^*_s=X^{\opi,\odel}_s(0,x)$.

By the BDG inequality, Proposition~\ref{P-v2-bound}, \ref{P-v1-v0} and \ref{P-x},
there exists a positive constant $C$ such that
\bea
&&\mbb{E}\left[\sup_{s\in[0,T]}\left|\int_0^s Z(u,X_{u}^*)dW_u\right|\right]
\leq C\mbb{E}\left[\left(\int_0^T |Z(s,X_s^*)|^2ds\right)^{\frac{1}{2}}\right]\nn \\
&& \leq C\mbb{E}\left[
\left(\int_0^T |Z_2(s)|^2|X_s^*|^4 ds\right)^{\frac{1}{2}}+
\left(\int_0^T |Z_1(s)|^2|X_s^*|^2 ds\right)^{\frac{1}{2}}+
\left(\int_0^T |Z_0(s)|^2ds\right)^{\frac{1}{2}}\right]\nn \\
&&\leq C\mbb{E}\left[1+||X^*||^4_T+\int_0^T
\Bigl(|Z_2(s)|^2+|Z_1(s)|^2+|Z_0(s)|^2\Bigr)ds\right]<\infty
\eea
where  in the second inequality, we have used the fact that
\be
(a+b+c)^{1/2}\leq \sqrt{a}+\sqrt{b}+\sqrt{c}~.
\ee
for every $a,b,c\geq 0$. Thus, $\Bigl(\int_0^sZ(u,X_u^*)dW_u\Bigr)_{s\in[0,T]}$ is 
a martingale.

For the integrations by the counting and marked point processes,
it is suffice to check that the integration by the corresponding compensator 
is in $\mbb{L}^1(\Omega)$ (See,  Corollary C4, Chapter VIII in \cite{Bremaud}.).
Therefore, for the second term, we need to check
\bea
\mbb{E}\left[\int_0^T\bigl|V(u,X_{u}^*+\del_u^*)-V(u,X_{u}^*)\bigr|\lambda_u du\right]<\infty
\eea
In fact,
\bea
&&\mbb{E}\left[\int_0^T\bigl|V(u,X_{u}^*+\del_u^*)-V(u,X_{u}^*)\bigr|\lambda_u du\right] \nn \\
&&=\mbb{E}\left[ \int_0^T \lambda_u\bigl|V_2(u)(2X_u^*\del_u^*+(\del_u^*)^2)+2V_1(u)\del_u^*\bigr|du\right]\nn \\
&&\leq C\mbb{E}\left[\int_0^T \Bigl(|X_u^*|^2+|\del_u^*|^2+|V_1(u)|^2\Bigr)du\right]<\infty.
\eea
Similarly, for the third term, 
\bea
&&\mbb{E}\left[\int_0^T\int_K \bigl| V(u,X_{u}^*+z)-V(u,X_u^*)\bigr|\Lambda(u,z)dudz\right]\nn \\
&&=\mbb{E}\left[\int_0^T \int_K \bigl| V_2(u)(2X_{u}^* z+z^2)+2V_1(u)z\bigr|\Lambda(u,z)dudz\right]\nn \\
&&\leq C\mbb{E}\left[\int_0^T\Bigl(|X_u^*|^2+|V_1(u)|^2+|\Phi_{2,u}|\Bigr)du\right] <\infty
\eea
where we have used the boundedness of the compensator and the support $K$.
The above facts combined with the construction of $a(t,x)$, strict positivity 
of $M$ as well as $\lambda[V_2+\eta]$, guarantee that the optimality principle in Proposition~\ref{P-optimality}
is indeed satisfied. 
\end{proof}
\end{theorem}

In Appendix A, an investigation of the relation between the penalty size and 
the remaining position at the terminal time is given.
It is proved that the terminal position size $X_T^*$ can be made arbitrary small by 
increasing the size of the penalty $\xi$.
This result implies that the proposed strategy can also be used for the 
liquidation problem in the presence of uncertain customer order flows.

\section{An extension to a  portfolio position management}
In the following sections, we are going to extend the previous framework 
so that we can deal with the optimal position management for a market maker
in the presence of 
$n\in\mbb{N}$ securities.
Firstly, Let us summarize the assumptions below. As before, the definition of
each variable will appear along the discussions in the following sections.

\subsubsection*{$\mathbf{Assumption~A^\prime}$} 
{\it 
$\caln^i(\omega, dt,dz), i\in\{1,\cdots,n\}$ is a random counting measure of a marked point process  
with a bounded support $K\subset\mbb{R}\backslash\{0\}$ for its mark $z$,
and $H^i, i\in\{1,\cdots,n\}$ is a counting process.
All the stochastic processes which do not jump by $\caln^i,~H^i$ for $i\in\{1,\cdots,n\}$ 
are assumed to be $\mbb{F}^W$-adapted
and hence continuous. This $\mbb{F}^W$ adaptedness 
includes all the stochastic processes defined below. \vspace{2mm}
\\
$(a_1^\prime)$ $\bS:\Omega\times[0,T]\rightarrow \mbb{R}^n$ is non-negative and $\bS\in\mbb{S}_n^4(0,T)$.\\
$(a_2^\prime)$ $\bbs, \bl: \Omega\times[0,T] \rightarrow \mbb{R}^n$ and $\bbs,\bl \in\mbb{S}_n^4(0,T)$. \\ 
$(a_3^\prime)$ $\Lambda^i(\cdot,\cdot): \Omega\times [0,T] \times \mbb{R} \rightarrow \mbb{R}$ 
for $i\in \{1,\cdots,n\}$ are such that
$\Lambda^i(t,\cdot)(\omega)$ is a non-negative measurable function with bounded support $K\subset \mbb{R}\backslash\{0\}$ 
for every $t\in[0,T]$ and $\omega\in\Omega$, 
and that $\Lambda^i(\cdot,z)$ is a uniformly bounded $\mbb{F}^W$-adapted process for every $z\in K$. \\
$(a_4^\prime)$ $\wt{\gamma}:\Omega\times[0,T]\rightarrow \mbb{R}^{n\times n}$ is uniformly bounded and takes 
values in the space of $n\times n$ symmetric positive-semidefinite matrices.\\
$(a_5^\prime)$ $M: \Omega\times [0,T] \rightarrow\mbb{R}^{n\times n}$ is uniformly bounded  and takes values 
in the space of $n\times n$ symmetric positive-definite matrices. \\
$(a_6^\prime)$ $\wt{\xi}:\Omega\rightarrow \mbb{R}^{n\times n}$ is bounded, $\calf_T^W$-measurable and takes values 
in the space of $n\times n$  symmetric positive-semidefinite matrices. \\
$(a_7^\prime)$ $\lambda^i, \wt{\eta}^i: \Omega\times[0,T]\rightarrow \mbb{R}$ for $i\in \{1,\cdots, n\}$
are uniformly bounded and strictly positive. \\
$(a_8^\prime)$ $\beta:\Omega \times[0,T]\rightarrow \mbb{R}^{n\times n}$ is uniformly bounded and 
takes values in the space of $n\times n$ symmetric matrices. \\
$(a_9^\prime)$ There is no simultaneous jump among $(\caln^i, H^i)_{i\in\{1,\cdots,n\}}$.
}

\subsection{The market description}
We consider a market quite similar to what is described in Section~\ref{sec-market}, but now with 
$n$ securities. The market maker's position for the securities starting $\bx\in\mbb{R}^n$ at time $t$
is given by the following $n$-dimensional vector process:
\bea
\bX_s^{\pi,\del}(t,\bx)=\bx+\sum_{i=1}^n\int_t^s\int_K \bee_i z\caln^i(du,dz)+\int_t^s \bpi_u du+
\sum_{i=1}^n\int_t^s \bee_i \del_u^i dH_u^i
\eea
where $\bee_i,~i\in\{1,\cdots,n\}$ is the unit $\mbb{R}^n$ vector whose elements are all zero 
except the $i$-th element given by $1$. $\bpi=(\pi^i)_{i\in \{1,\cdots,n\}}$
and $\bdel=(\del^i)_{i\in\{1,\cdots,n\}}$ are  $\mbb{F}$-predictable trading strategies of the market maker 
in the exchange and in the dark pool, respectively.  The superscript $i$ is added to distinguish 
the corresponding security. $H^i, i\in\{1,\cdots,n\}$ is the counting process denoting the 
occurrence of the execution of the $i$-th security in the dark pool.
$\caln^i, i\in\{1,\cdots,n\}$ is the counting measure which describes the occurrence of an incoming customer order of the $i$-th security and its size.  

We suppose that there exists a common bounded support $K\subset \mbb{R}\backslash\{0\}$ for the size of the
incoming orders. We assume as before the existence of the compensators such that
\bea
&&\int_0^t\int_K \wt{\caln}^i(ds,dz)= \int_0^t \int_K \Bigl(\caln^i(ds,dz)-\Lambda^i(s,z)dzds\Bigr) \nn \\
&&\int_0^t d\wt{H}_s^i=\int_0^t \Bigl(dH_s^i-\lambda_s^ids\Bigr)
\eea
for $t\in[0,T]$ are $\mbb{F}$-martingales for every $i\in\{1,\cdots,n\}$. Let us also set
the stochastic processes $\bPhi=(\Phi^i)_{i\in \{1,\cdots,n\}}$, $\bPsi=(\Psi^i)_{i\in \{1,\cdots,n\}}$
and $\bPhi_2=(\Phi^i_2)_{i\in \{1,\cdots,n\}}$ representing the moments of the order size by
\bea
\Phi^i_t:=\int_K z\Lambda^i(t,z)dz,\quad \Psi_t^i:=\int_K |z|\Lambda^i(t,z)dz,\quad 
\Phi_{2,t}^i:=\int_K z^2 \Lambda^i(t,z)dz~,
\eea
for $t\in[0,T]$ which are all uniformly bounded by Assumption $A^\prime$.

We assume that the price vector $\wt{\bS}^{\pi,\del}(t,\bx)=\bigl(S_i^{\pi,\del}(t,\bx)\bigr)_{i\in\{1,\cdots,n\}}$,
which denotes the market price observed in the exchange under the impact of the 
market maker's strategy $(\bpi,\bdel)$ starting from the position size $\bx$ at time $t$, 
is given by
\bea
\wt{\bS}^{\pi,\del}_s(t,\bx)=\bS_s+M_s\bpi_s-\beta_s \bX_s^{\pi,\del}(t,\bx)~
\eea
for $s\in[t,T]$.
Here, $M$ and $\beta$ are not necessarily diagonal and hence they can induce direct as well as 
contagious stochastic linear price impacts from the continuous trading and also from
the aggregate reactions of the other investors regarding the inventory size of the market maker.
We can naturally imagine that, for example, due to the proxy hedging by correlated assets,  
a high trading speed or a big outstanding position of a certain security  induces 
similar price actions among the closely related assets.
\subsection{The market maker's problem}
We model the cash flow in the interval $]t,T]$ to the market maker with strategy $(\bpi,\bdel)$ as
\bea
&&-\int_t^T \wt{\bS}_s^{\pi,\del}(t,\bx)^\top \bpi_s ds
-\sum_{i=1}^n \int_t^T \int_K \wt{S}^{\pi,\del}_{i,s-}(t,\bx)(1-{\rm sgn}(z)b_s^i)z\caln^i(ds,dz) \nn \\
&&\quad-\sum_{i=1}^n \int_t^T \Bigl(\bigl(\bS_s-\beta_s \bX_{s-}^{\pi,\del}(t,\bx)\bigr)^i \del^i_s+\wt{\eta}_s^i|\del^i_s|^2
\Bigr)dH_s^i+\int_t^T \bl^\top_s \bX_s^{\pi,\del}(t,x)ds,
\eea
where the symbol $\top$ denotes the transposition.
We consider the following market maker's problem:
\bea
&&\wt{V}(t,\bx)={\rm ess}\inf_{\bpi,\bdel\in \calu} \mbb{E}\left[\LDis \right.
(\bX_T^{\pi,\del})^\top \wt{\xi} \bX_T^{\pi,\del} + \int_t^T (\bX_s^{\pi,\del})^\top \wt{\gamma}_s
\bX_s^{\pi,\del} ds \nn \\
&&+\int_t^T\bigl( (\wt{\bS}_s^{\pi,\del})^\top \bpi_s-\bl_s^\top \bX_s^{\pi,\del}\bigr)ds
+\sum_{i=1}^n \int_t^T \int_K \wt{S}^{\pi,\del}_{i,s-}\bigl(1-{\rm sgn}(z)b^i_s\bigr)z\caln^i(ds,dz)\nn \\
&&+\sum_{i=1}^n \int_t^T \Bigl( \bigl(\bS_s-\beta_s\bX_{s-}^{\pi,\del}\bigr)^i\del_s^i+\wt{\eta}_s^i|\del_s^i|^2
\Bigr)dH_s^i \left.\LDis \Bigr|\calf_t \right],
\label{mm-problem-multi}
\eea
where we have omitted the argument $(t,\bx)$  to save the space. 
By making $\wt{\xi}$ and $\wt{\gamma}$ proportional to the (stochastic) covariance matrix among 
the securities, the market maker can include the portfolio diversification effects.
In the above modeling, the customer orders and the executions in the dark pool
are assumed to occur independently for each security. However, it is not difficult to introduce 
simultaneous customer orders or the dark pool executions for an arbitrary subset of the securities
by following the idea of dynamic Markov copula model studied by Bielecki, Cousin, 
Cr\'epey \& Herbertsson~(2014a, 2014b)~\cite{Bielecki-2014a,Bielecki-2014b}.
If there exist strong clusterings among the customer orders or the executions,
an extension to this direction may become worthwhile.

The set of admissible strategies $\calu$ is defined below.
\begin{definition}
We define the admissible strategies $\calu$ by the set of $\mbb{F}$-predictable processes $(\bpi,\bdel)$
that belong to $\mbb{H}_n^2(0,T)\times \mbb{H}_n^2(0,T)$ and also Markovian with respect to the position size,
i.e., they are expressed with some measurable functions $(f^\pi, f^\del)$ by
\bea
\bpi_s=f^\pi(s,\bX_{s-}^{\pi,\del}(t,\bx)), \quad \bdel_s=f^{\del}(s,\bX_{s-}^{\pi,\del}(t,\bx))
\label{D-U-markov2}
\eea
where, for $a\in\{\pi,\del\}$, $f^a: \Omega \times [0,T] \times \mbb{R}^n \rightarrow \mbb{R}^n$ 
and  $f^{a}(\cdot,\bx)$ is an $\mbb{F}^W$-adapted process for all $\bx\in\mbb{R}^n$. 
\end{definition}

Let us write the dynamics of the $\mbb{F}^W$-adapted bounded process $\beta$ as
\bea
d\beta_t=\mu_t^\beta dt+\sum_{j=1}^d \bigl(\sigma_t^\beta\bigr)_j dW^j_t
\eea
and define
\bea
&&\xi:=\wt{\xi}-\frac{\beta_T}{2}, \quad \gamma:=\wt{\gamma}+\frac{\mu^\beta}{2}~ \nn \\
&&\eta^i:=\wt{\eta}^i+\frac{(\beta)_{i,i}}{2},~~{\rm for}~~ i\in\{1,\cdots,n\}~.
\eea
\subsubsection*{$\bold{Assumption~B^\prime}$}
{\it
$(b_1^\prime)$ $\mu^\beta: \Omega\times[0,T]\rightarrow \mbb{R}^{n\times n}$ and 
$(\sigma^\beta)_i,i\in\{1,\cdots,d\}: \Omega\times[0,T]\rightarrow \mbb{R}^{n\times n}$ are uniformly bounded, 
symmetric and $\mbb{F}^W$-adapted. \\
$(b_2^\prime)$ $\xi$ is positive-semidefinite. \\
$(b_3^\prime)$ $\gamma$ is positive-semidefinite $d\mbb{P}\otimes dt$-a.e..\\
$(b_4^\prime)$ There exists a constant $c>0$ such that 
 $\lambda^i \eta^i\geq c~d\mbb{P}\otimes dt$-a.e. for every $i\in\{1,\cdots, n\}$
 and also $\by^\top M \by \geq c|\by|^2~ d\mbb{P}\otimes dt$-a.e. for every $\by\in \mbb{R}^n$.
}

\begin{definition}
The cost function for the market maker with a given position size $\bx\in \mbb{R}^n$ at $t\in[0,T]$ is
\bea
&&J^{t,\bx}(\bpi,\bdel)=\mbb{E}\left[\LDis \right.  (\bX_T^{\pi,\del})^\top \xi \bX_T^{\pi,\del}
+\int_t^T \Bigl( (\bX_s^{\pi,\del})^\top \gamma_s \bX_s^{\pi,\del}
+(\bX_s^{\pi,\del})^\top \bigl(\beta_s (\bbs\bPsi)_s-\bl_s\bigr) \Bigr)ds\nn \\
&&\hspace{-5mm}+\left.\int_t^T \left\{ \bpi_s^\top M_s \bpi_s+
\bigl(\bS_s+M_s\bTheta_s)^\top \bpi_s
+\bS_s^\top \bTheta_s+\sum_{i=1}^n \Bigl(\lambda_s^i\bigl[\eta^i_s (\del_s^i)^2+S_s^i\del_s^i\bigr]+ \frac{(\beta_s)_{i,i}}{2}\Phi_{2,s}^i\Bigr)\right\}ds \Bigr|\calf_t^W\right]\nn \\
\eea 
where $\bTheta,~\bbs\bPsi: \Omega\times[0,T]\rightarrow \mbb{R}^n$ are defined by $(\bTheta_s)^i:=\Phi^i_s-b_s^i\Psi^i_s$ and $(\bbs\bPsi)_s^i=b_s^i\Psi_s^i$ for $i\in\{1,\cdots,n\}$.
Here, the argument $(t,\bx)$ of the position size is omitted to save the space.
\end{definition}

\begin{proposition}
Under Assumptions $A^\prime$ and $B^\prime$, the market maker's problem (\ref{mm-problem-multi}) is equivalent to
\be
V(t,\bx)={\rm ess}\inf_{(\bpi,\bdel)\in\calu}J^{t,\bx}(\bpi,\bdel)
\label{eq-mm-problem-multi}
\ee
and it has a unique optimal solution $(\bpi^*,\bdel^*)\in \calu$.
\label{P-mm-problem-multi}
\begin{proof}
By using
\bea
&&-\int_t^T (\bX_{s-}^{\pi,\del})^\top \beta_sd\bX_s^{\pi,\del}=-\frac{1}{2}(\bX_T^{\pi,\del})^\top\beta_T \bX_T^{\pi,\del}
+\frac{1}{2}\bx^\top \beta_t \bx\nn \\
&&\qquad\qquad+\int_t^T\left(\frac{1}{2}(\bX_s^{\pi,\del})^\top\mu^\beta_s\bX_s^{\pi,\del}ds
+\frac{1}{2}(\bX_s^{\pi,\del})^\top \sigma^\beta_s \bX_s^{\pi,\del}\cdot dW_s\right)\nn \\
&&\qquad\qquad+\sum_{i=1}^n \left(\int_t^T \frac{1}{2}(\beta_s)_{i,i}(\del_s^i)^2dH_s^i
+\int_t^T\int_K\frac{1}{2}(\beta_s)_{i,i}z^2\caln^i(ds,dz)\right)\nn 
\eea
and redefining the value function
\bea
V(t,\bx):=\wt{V}(t,\bx)-\frac{1}{2}\bx^\top \beta_t \bx
\eea
one can prove it in exactly the same way as Proposition~\ref{P-mm-problem}.
\end{proof}
\end{proposition}

\section{Solving the problem with multiple securities}
\subsection{A candidate solution}
We derive a candidate solution for the market maker's problem.
Firstly, let us rewrite the optimality principle for the problem with multiple securities.

\begin{proposition} (Optimality Principle)
Let Assumptions $A^\prime$ and $B^\prime$ are satisfied. Then,\\
(a) For all $\bx\in\mbb{R}^n$, $(\bpi,\bdel)\in\calu$ and $t\in[0,T]$, the process
\bea
&&\left(\LDis \right.  V(s,\bX_s^{\pi,\del})+\int_t^s \Bigl( (\bX_u^{\pi,\del})^\top \gamma_u
\bX_u^{\pi,\del}+(\bX_u^{\pi,\del})^\top\bigl(\beta_u(\bbs\bPsi)_u-\bl_u\bigr)\Bigr)du +\int_t^s \left\{ \bpi_u^\top M_u 
\bpi_u \LDis \right. \nn \\
&&\left.\left.+\bigl(\bS_u+M_u\bTheta_u)^\top \bpi_u +\bS_u^\top \bTheta_u+\sum_{i=1}^n \Bigl(\lambda_u^i\bigl[\eta^i_u (\del_u^i)^2+S_u^i\del_u^i\bigr]+ \frac{(\beta_u)_{i,i}}{2}\Phi_{2,u}^i\Bigr)\right\}du\right)_{s\in[t,T]}
\eea
is an $\mbb{F}$-submartingale. \\
(b) $(\bpi^*,\bdel^*)$ is optimal if and only if 
\bea
&&\hspace{-10mm}\left(\LDis \right.  V(s,\bX_s^{\opi,\odel})+\int_t^s \Bigl( (\bX_u^{\opi,\odel})^\top \gamma_u
\bX_u^{\opi,\odel}+(\bX_u^{\opi,\odel})^\top\bigl(\beta_u(\bbs\bPsi)_u-\bl_u\bigr)\Bigr)du +\int_t^s \left\{ 
(\bpi_u^*)^\top M_u 
(\bpi_u^*) \LDis \right. \nn \\
&&\left.\left.+\bigl(\bS_u+M_u\bTheta_u)^\top \bpi_u^* +\bS_u^\top \bTheta_u+\sum_{i=1}^n \Bigl(\lambda_u^i\bigl[\eta^i_u (\odel_u^i)^2+S_u^i\odel_u^i\bigr]+ \frac{(\beta_u)_{i,i}}{2}\Phi_{2,u}^i\Bigr)\right\}du\right)_{s\in[t,T]}
\eea
is an $\mbb{F}$-martingale.
\end{proposition}

Derivation of a candidate solution and the associated stochastic HJB equation is similar to 
the single security case. We assume that the $\mbb{F}^W$ semimartingale $\Bigl(V(t,\bx)\Bigr)_{t\in[0,T]}$ has the 
following decomposition:
\bea
V(s,\bx)=V(t,\bx)+\int_t^s a(u,\bx)du+\int_t^s Z(u,\bx)dW_u
\eea
where $a: \Omega\times[0,T]\times \mbb{R}^n\rightarrow \mbb{R}$,
$Z:\Omega\times[0,T]\times \mbb{R}^n\rightarrow \mbb{R}^d$ and $a(\cdot,\bx)$ as well as
$Z(\cdot,\bx)$ are $\mbb{F}^W$-adapted processes for all $\bx\in\mbb{R}^n$.
We suppose that the value function can be decomposed, for every $t\in[0,T]$ and $\bx\in\mbb{R}^n$ as
\bea
V(t,\bx)=\bx^\top V_2(t) \bx+2x^\top V_1(t)+V_0(t) \\
Z(t,\bx)=\bx^\top Z_2(t) \bx+2x^\top Z_1(t)+Z_0(t)
\eea
where $V_2:\Omega\times[0,T]\rightarrow \mbb{R}^{n\times n}$, 
$V_1:\Omega\times[0,T]\rightarrow \mbb{R}^n$, $V_0:\Omega\times[0,T]\rightarrow \mbb{R}$,
$Z_2:\Omega\times[0,T]\rightarrow \mbb{R}^{n\times n\times d}$, $Z_1:\Omega\times[0,T]\rightarrow \mbb{R}^{n\times d}$
and $Z_0:\Omega\times [0,T]\rightarrow \mbb{R}^d$ are 
all $\mbb{F}^W$-adapted processes. 
In addition, $V_2$ and $Z_2$ (with respect to the first two indexes)
are symmetric. 

A lengthy but straightforward calculation shows that
a necessary condition for the optimality principle is
\bea
&&a(u,\bx)+\bx^\top \gamma_u \bx+\bx^\top (\beta_u(\bbs\bPsi)_u-\bl_u)+\bS_u^\top \bTheta_u+
\sum_{i=1}^n\frac{(\beta_u)_{i,i}}{2}\Phi_{2,u}^i\nn \\
&&+2\bx^\top V_2(u)\bPhi_u+2V_1(u)^\top \bPhi_u+\sum_{i=1}^n[V_2(u)]_{i,i}\Phi_{2,u}^i\nn 
\eea
\vspace{-5mm}
\bea
&&\hspace{-10mm}+\inf_{\bpi,\bdel}\left\{\LDis \right. \Bigl(\bpi+M_u^{-1}\bigl[V_2(u)\bx+V_1(u)+\frac{1}{2}(\bS_u+M_u\bTheta_u)\bigr]
\Bigr)^\top M_u \nn \\
&&\hspace{10mm} \times \Bigl(\bpi+M_u^{-1}\bigl[V_2(u)\bx+V_1(u)+\frac{1}{2}(\bS_u+M_u\bTheta_u)\bigr]
\Bigr)\nn \\
&&+\sum_{i=1}^n \lambda^i_u\bigl([V_2(u)]_{i,i}+\eta_u^i\bigr)
\Bigl(\del^i+\frac{\bigl[V_2(u)\bx+V_1(u)+\frac{1}{2}\bS_u\bigr]^\top \bee_i}{[V_2(u)]_{i,i}+\eta_u^i}\Bigr)^2\nn 
\eea
\vspace{-4mm}
\bea
&&-\Bigl[V_2(u)\bx+V_1(u)+\frac{1}{2}(\bS_u+M_u\bTheta_u)\Bigr]^\top M_u^{-1}\Bigl[V_2(u)\bx+V_1(u)+\frac{1}{2}(\bS_u+M_u\bTheta_u)\Bigr]\nn \\
&&\left.\LDis-\sum_{i=1}^n \lambda^i_u\frac{\displaystyle \Bigl(\bigl[V_2(u)\bx+V_1(u)+\frac{\bS_u}{2}\bigr]^\top \bee_i\Bigr)^2}{[V_2(u)]_{i,i}+\eta^i_u}\right\}=0\quad d\mbb{P}\otimes dt-a.e.,
\eea
where we need $[V_2]_{i,i}+\eta^i>0$ $d\mbb{P}\otimes dt$-a.e. for every $i\in\{1,\cdots,n\}$.
As a result, we obtain the following.

\subsubsection*{A Candidate Solution}
{\it
A ``candidate" of the optimal solution and the
corresponding value function for the market maker's problem (\ref{eq-mm-problem-multi})
are given by
\bea
\label{eq-opimulti}
&&\pi_u^*=-M_u^{-1}\Bigl(V_2(u)\bX_{u-}^{\opi,\odel}(t,\bx)+V_1(u)+\frac{1}{2}\bigl(\bS_u+M_u\bTheta_u\bigr)\Bigr)\\
&&(\del_u^*)^i=-\frac{\displaystyle \bigl[V_2(u)\bX_{u-}^{\opi,\odel}(t,\bx)+V_1(u)+\frac{1}{2}\bS_u\bigr]^i}{[V_2(u)]_{i,i}+\eta_u^i},\quad{\rm for}\quad i\in\{1,\cdots,n\}
\label{eq-odelmulti}
\eea
for $u\in[t,T]$ and $V(t,\bx)=\bx^\top V_2(t)\bx+2\bx^\top V_1(t)+V_0(t)$, respectively.
Here, $\bX^{\opi,\odel}(t,\bx)$ is the solution of 
\bea
X_s^{\opi,\odel}(t,\bx)=\bx+\sum_{i=1}^n\int_t^s\int_K \bee_i z \caln^i(du,dz)+
\int_t^s \bpi_u^* du+\sum_{i=1}^n \int_t^s \bee_i (\del_u^*)^i dH_u^i, ~s\in[t,T]
\label{eq-xopt-multi}
\eea
$(V_2,Z_2)$, $(V_1,Z_1)$ and $(V_0,Z_0)$ must be the well-defined solutions
of the following three BSDEs
\bea
\label{eq-v2-multi}
&&V_2(t)=\xi+\int_t^T\left\{-V_2(u)\left[M_u^{-1}+{\rm diag}\Bigl(
\frac{\lambda_u}{V_2(u)+\eta_u}\Bigr)\right]V_2(u)+\gamma_u\right\}du-\int_t^T Z_2(u) dW_u\nn \\ 
\eea
\vspace{-10mm}
\bea
\label{eq-v1-multi}
&&V_1(t)=-\int_t^T\left\{\LDis \right. V_2(u)\left[M_u^{-1}+{\rm diag}\Bigl(\frac{\lambda_u}{V_2(u)+\eta_u}\Bigr)
\right]V_1(u)-\frac{1}{2}\bigl(\beta_u(\bbs\bPsi)_u-\bl_u\bigr)\nn \\
&&\hspace{5mm}\left. \LDis +V_2(u)\left(\left[M_u^{-1}+{\rm diag}\Bigl(
\frac{\lambda_u}{V_2(u)+\eta_u}\Bigr)\right]\frac{\bS_u}{2}-\frac{1}{2}\bTheta_u-(\bbs\bPsi)_u\right)\right\}du
-\int_t^T Z_1(u)dW_u\nn \\ 
\eea
\bea
\label{eq-v0-multi}
&&V_0(t)=-\int_t^T \left\{\LDis \right. \Bigl(V_1(u)+\frac{\bS_u}{2}\Bigr)^\top 
\left[M_u^{-1}+{\rm diag}\Bigl(\frac{\lambda_u}{V_2(u)+\eta_u}\Bigr)\right]\Bigl(V_1(u)+\frac{\bS_u}{2}\Bigr)\nn \\
&&\hspace{20mm}-\bigl(\bPhi_u+(\bbs\bPsi)_u\bigr)^\top V_1(u)-\sum_{i=1}^n[V_2(u)]_{i,i}\Phi^i_{2,u}\nn \\
&&\hspace{10mm}\left.\LDis-\frac{1}{2}\bigl(\bS_u^\top \bTheta_u+\sum_{i=1}^n[\beta_u]_{i,i}\Phi^i_{2,u}\bigr)+
\frac{1}{4}\bTheta_u^\top M_u \bTheta_u \right\}du-\int_t^T Z_0(u)dW_u
\eea
satisfying, for every $i\in\{1,\cdots,n\}$, 
\bea
[V_2]_{i,i}+\eta^i>0 
\eea
$d\mbb{P}\otimes dt$-a.e. in $\Omega\times[0,T]$. In the above, ${\rm diag}\Bigl(\frac{\lambda_u}{V_2(u)+\eta_u}\Bigr)$
is defined as a diagonal matrix whose $(i,i)$-th element $i\in\{1,\cdots,n\}$ is given by
$\displaystyle \frac{\lambda_u^i}{[V_2(u)]_{i,i}+\eta^i_u}$.
}

\subsection{Verification}
In the multiple-security setup, $V_2$ follows a non-linear matrix valued 
BSDE. Since there is no comparison theorem known for a multi-dimensional BSDE
in general, we cannot apply the technique used in the single-security case.
Interestingly however, we shall see $V_2$ is the backward stochastic Riccati 
differential equation (BSRDE) associated with a special type of stochastic 
linear quadratic control (SLQC) problem in a diffusion setup studied by 
Bismut (1976)~\cite{Bismut}.

\begin{theorem}
\label{T-v2-multi}
Under Assumptions $A^\prime$ and $B^\prime$, there exists a unique solution of
$(V_2,Z_2)$ for the BSDE (\ref{eq-v2-multi}). In particular, $V_2$ takes values in the space of $n\times n$ symmetric 
positive-semidefinite matrices and is a.s. uniformly 
bounded i.e., there exists a positive constant $C^\prime$ such that
\be
{\rm ess}\sup \left(\sup_{t\in[0,T]}\bigl|V_2(t)\bigr|(\omega)\right)\leq C^\prime~,
\ee
and $Z_2\in \mbb{H}^p_{n\times n\times d}(0,T)$ for any $p>0$.
\begin{proof}
Let us introduce an $n$-dimensional Brownian motion $w$ which is orthogonal to $W$
and consider $\calf_t^\prime:=\calf_t^W\vee \calf_t^w$
where $\calf_t^w$ is the augmented filtration generated by $w$.
We study an $n$-dimensional $\bigl(\mbb{F}^\prime:=(\calf_t^\prime)_{t\geq 0}\bigr)$-adapted vector process
staring from $\bx\in\mbb{R}^n$ at time $t$ which is 
controlled by the $2n$-dimensional vector process $\btheta$:
\bea
\bX_s^{\theta}(t,\bx)=\bx+\int_t^s C_u\btheta_u du+
\sum_{j=1}^n \int_t^s D_u^j \btheta_u dw_u^j~,~s\in[t,T]~.
\eea
Here, $C:\Omega \times[0,T]\rightarrow \mbb{R}^{n\times 2n}$ is defined by
\bea
C_u:=\begin{pmatrix} \mbb{I}_{n\times n} & {\rm diag}(\lambda_u^i)
\end{pmatrix}
\eea
for $u\in[0,T]$, where $\mbb{I}_{n\times n}$ is the $n$-dimensional identity matrix, and ${\rm diag}(\lambda^i)$
is the $n$-dimensional diagonal matrix whose $(i,i)$-th element $i\in\{1,\cdots,n\}$ is given by $\lambda^i$.
We use the same notation for the diagonal matrices below.
$D^i:\Omega\times [0,T]\rightarrow \mbb{R}^{n\times 2n}$ for $i\in\{1,\cdots,n\}$
has zero entry for all  except the $(i,n+i)$-th element which is given  by
\be
[D_u^i]_{i,n+i}=\sqrt{\lambda^i_u}~
\ee
for $u\in[0,T]$.
We define the admissible strategies $\calu^\prime$
as the set of $2n$-dimensional $\mbb{F}^\prime$-adapted processes $\btheta$
that belong to $\mbb{H}^2_{2n}(0,T)$. 
\\

Now, let us consider the following SLQC problem:
\bea
V^\prime(t,\bx)={\rm ess}\inf_{\btheta\in \calu^\prime}\mbb{E}
\left[\LDis  (\bX_t^{\theta})^\top \xi \bX_T^\theta
+\int_t^T \Bigl((\bX_s^{\theta})^\top \gamma_s \bX_s^{\theta}
+\btheta_s^\top N_s \btheta_s\Bigr)ds\Bigr|\calf_t^\prime\right]
\eea
where the argument $(t,\bx)$ is omitted from $\bX$ to save the space, and 
$N:\Omega\times[0,T]\rightarrow \mbb{R}^{2n\times 2n}$ is defined for $u\in[0,T]$ by
\bea
N_u=\begin{pmatrix} M_u & \bvec{0}_{n\times n} \\
\bvec{0}_{n\times n} & {\rm diag} (\lambda^i_u \eta_u^i)~
\end{pmatrix}.
\eea

Then, by Proposition 5.1 in \cite{Bismut}, 
the associated BSRDE is given by
\be
P(t)=\xi+\int_t^T \left\{-P(u)C_u \Bigl(N_u+\sum_{i=1}^n (D^i_u)^\top P(u) D^i_u\Bigr)^{-1} C_u^\top P(u)
+\gamma_u\right\}du-\int_t^T Z_P(u)dW_u
\ee
where $P$ is connected to the value function as $V^\prime(t,\bx)=\bx^\top P(t) \bx$.
Note that the stochastic integration by $dw$ vanishes because $W\perp w$ and that
the terminal value $\xi$ and all the processes included in the driver are 
$\mbb{F}^W$-adapted. 
By noticing that
\bea
N_u+\sum_{i=1}^n (D^i_u)^\top P(u) D^i_u=\begin{pmatrix} M_u & \bvec{0}_{n\times n} \\
\bvec{0}_{n\times n} & {\rm diag}\Bigl(\lambda^i_u ([P(u)]_{i,i}+\eta^i_u)\Bigr) \end{pmatrix}
\label{eq-nhat}
\eea
one can confirm that the BSDE of $P$ is equal to that of $V_2$ given by 
$(\ref{eq-v2-multi})$~\footnote{It is not difficult to confirm the same BSRDE arises 
as the stochastic HJB equation by the same method we used.}.

Under Assumptions $A^\prime$ and $B^\prime$, $\xi$
is positive-semidefinite and bounded, $\gamma$ is positive-semidefinite and uniformly bounded,
$C$, $D$ and $N$ are uniformly bounded. In particular, 
there exists a constant $c>0$ such that
\bea
\bvec{y}^\top N_u \bvec{y} \geq c|y|^2, \quad d\mbb{P}\otimes dt-a.e.
\eea
for all $y\in\mbb{R}^{2n}$.  Thus, by Theorem 6.1 in \cite{Bismut}, $P$ (and hence $V_2$)
has a unique solution, which is symmetric, positive-semidefinite and a.s. uniformly bounded.
In particular, this implies $[P(u)]_{i,i}\geq 0,~~d\mbb{P}\otimes dt$-a.e.. 

Since $P$ is positive, one sees from (\ref{eq-nhat}),
\bea
0< \by^\top \Bigl(N_u+\sum_{i=1}^n(D^i_u)^\top P(u) D_u^i\Bigr)^{-1} \by \leq \frac{|\by|^2}{c}
\quad d\mbb{P}\otimes dt-a.e.
\eea
 for all $\by\in \mbb{R}^{2n}$ and hence $\Bigl(N_u+\sum_{i=1}^n(D^i_u)^\top P(u) D_u^i\Bigr)^{-1}_{u\in[0,T]}$
is a uniformly bounded linear operator.
Using the boundedness of $P$ and the other variables, one sees
\bea
m_t=P(t)-P(0)-\int_0^t\left\{P(u)C_u\Bigl(N_u+\sum_{i=1}^n (D_u^i)^\top P(u) D_u^i\Bigr)^{-1}C_u^\top P(u)-\gamma_u
\right\}du
\eea
for $t\in[0,T]$ is a uniformly bounded martingale. Thus, from the BDG inequality, for any $p>0$, 
there exists a positive constant $C$ such that
\bea
\mbb{E}\left[\Bigl(\int_0^T |Z_P(u)|^2 du\Bigr)^{p/2}\right]\leq C\mbb{E}\Bigl[||m||_T^p\Bigr]<\infty
\eea
and hence $Z_P$ (and so does $Z_2$) belongs to $\mbb{H}^p_{n\times n\times d}(0,T)$ for $\forall p>0$.
\end{proof}
\end{theorem}
For more general results on the SLQC problem and the associated BSRDE,
we refer to Peng~(1992)~\cite{Peng} and Tang~(2003, 2014)~\cite{Tang-2003,Tang-2014},
where the assumption of the orthogonality ``$w\perp W$" is removed.

The following results are obtained in exactly the same way in Proposition~\ref{P-v1-v0}, \ref{P-x} and
Corollary~\ref{C-opt}.
\begin{proposition}
Under Assumptions $A^\prime$ and $B^\prime$, there exist unique solutions 
$(V_1,Z_1)\in \mbb{S}^4_n(0,T)\times \mbb{H}^4_{n\times d}(0,T)$ for (\ref{eq-v1-multi}),
and $(V_0,Z_0)\in \mbb{S}^2(0,T)\times \mbb{H}^2_d(0,T)$ for (\ref{eq-v0-multi}), respectively.
Furthermore, the process for the position size $\Bigl(\bX_s^{\opi,\odel}(t,\bx)\Bigr)_{s\in[t,T]}$
given by (\ref{eq-xopt-multi}) belongs to $\mbb{S}^4_n(t,T)$. 
The candidate solution $(\opi,\odel)$ given by (\ref{eq-opimulti})
and (\ref{eq-odelmulti}) is well-defined 
and satisfies $(\opi,\odel)\in \mbb{S}^4_n(t,T)\times \mbb{S}^4_n(t,T)\subset \calu$.
\end{proposition}

The above results establish the main theorem.
\begin{theorem}
Under Assumptions $A^\prime$ and $B^\prime$, the candidate solution $(\opi,\odel)$ given by (\ref{eq-opimulti})
and (\ref{eq-odelmulti})
is, in fact, the unique optimal solution of the market maker's problem given by (\ref{eq-mm-problem-multi}).
\begin{proof}
The proof is the same as that of Theorem~\ref{T-main}.
\end{proof}
\end{theorem}

\subsubsection*{Remark: A determination of the bid/offer spreads}
Before closing the section, let us comment on a possible determination 
of the bid/offer spreads $\bbs$.
Although we have assumed that the market maker do not dynamically control 
the bid/offer spreads to give a bias to the order flows, 
it is important of course to use a sustainable spread size for its market making business.
Suppose, for example, the spread size $b^i$  is proportional
to the volatility $|\sigma^i|$ of the $i$-th security as
\be
b^i_s= \hat{a} |\sigma_s^i|
\ee
where $\hat{a}>0$ is some constant and $i\in\{1,\cdots,n\}$.
Even if the intensity (and/or distribution) of the customer orders
is a non-linear function of $(b^i)_{i\in\{1,\cdots,n\}}$, 
the market maker can obtain the cost function or the distribution of its revenue 
by running the simulation based on the optimal strategy $(\bpi^*,\bdel^*)$ for each choice of $\hat{a}$,
which will give enough information to fix the size of $\hat{a}$.
\section{Implementation for a simple case}
In this section, we discuss the 
evaluation scheme for a simple case where $V_2$ becomes 
non-random. As we shall see below, the implementation of 
the optimal strategy is quite simple in this case.

Consider a setup where
$\xi,\gamma, M,\eta, \lambda$ (and hence naturally so is $\beta$) 
are non-random.
In this case, $V_2$ is a solution of the following  matrix-valued ordinary differential equation (ODE):
\bea
&&\frac{dV_2(s)}{ds}=V_2(s)\left(M_s^{-1}+{\rm diag}\Bigl(\frac{\lambda_s}{V_2(s)+\eta_s}\Bigr)\right)V_2(s)-\gamma_s,
\quad s\in[t,T]  \\
&&V_2(T)=\xi
\eea
which is the same ODE studied by Kratz \& Sch\"oneborn~\cite{Kratz}. 
As long as $\xi,\gamma, M,\eta, \lambda$ satisfy the boundedness conditions in Assumptions $A^\prime$ and $B^\prime$,
this Riccati equation has a positive bounded solution.
It is not difficult to numerically solve this equation 
by the standard technique for ODEs. In contrast to the model in \cite{Kratz}, we still need to evaluate $V_1$
to implement the optimal strategy (See Eqs (\ref{eq-opimulti}) and (\ref{eq-odelmulti}).).

For notational simplicity, let us put
\bea
F(s):=V_2(s)\left(M_s^{-1}+{\rm diag}\Bigl(\frac{\lambda_s}{V_2(s)+\eta_s}\Bigr)\right),~\quad s\in[t,T],
\eea
which is a deterministic matrix process. Let us also consider another
deterministic matrix process $Y_{t,\cdot}$ defined by the ODE
\bea
\frac{dY_{t,s}}{ds}=F(s)Y_{t,s}, \quad s\in[t,T]
\eea
where $Y_{t,t}=\mbb{I}_{n\times n}$.  Then, we have
\be
\frac{dY_{t,s}^{-1}}{ds}=-Y_{t,s}^{-1}F(s), \quad s\in[t,T]
\ee
with $Y_{t,t}^{-1}=\mbb{I}_{n\times n}$ and it is straightforward to obtain
\bea
V_1(s)=-Y_{t,s} \int_s^T Y_{t,u}^{-1} \mbb{E}
\left[ \frac{1}{2}F(u)\bS_u-
V_2(u)\Bigl(\frac{1}{2}\bTheta_u+(\bbs\bPsi)_u\Bigr)-\frac{1}{2}\bigl( \beta_u (\bbs\bPsi)_u-\bl_u\bigr)\Bigr|\calf_s^W
\right]du \nn \\
\eea
for $s\in[t,T]$. One sees that the bid/offer spread, the customer order flows
and the size of repo rate impact the optimal strategy though $V_1$.
Its evaluation only requires 
\bea
\mbb{E}\bigl[\bS_u|\calf_s^W\bigr], ~\mbb{E}\bigl[\bPhi_u|\calf_s^W\bigr], ~\mbb{E}\bigl[ (\bbs\bPsi)_u|\calf_s^W\bigr],~
\mbb{E}\bigl[\bl_u|\calf_s^W\bigr]~.
\eea
These quantities can be obtained analytically for simple models. Otherwise, one can 
apply the standard {\it small-diffusion} asymptotic expansion technique, which is
developed by Yoshida (1992a)~\cite{YoshidaE}, Takahashi (1999)~\cite{TA1}, Kunitomo \& Takahashi (2003)~\cite{KT2} for the 
pricing of European contingent claims, and also Yoshida (1992b)~\cite{Yoshida} for statistical applications. 
See Takahashi (2015)~\cite{T} and references therein for the recent developments.

\section{Implementation for a general Markovian case}
Although it is impossible to solve $V_2$ analytically in a general setup, 
getting an explicit expression of its approximation is very important for successful implementation 
of the proposed scheme. A similar BSDE 
is also relevant for solving a different type of optimal liquidation problem treated in \cite{Jeanblanc}.
Furthermore, considering the wide spread applications of SLQC problems in various engineering issues,  
developing a successful approximation scheme for a general BSRDE should be a very important 
research topic in its own light.

There exists an analytical approximation technique for non-linear BSDEs,
which was proposed in  Fujii \& Takahashi (2012a)~\cite{FT-12a}. 
The method introduces a perturbation to the driver and then 
linearizing the BSDE in each approximation order. It then adopts 
the small-diffusion asymptotic expansion to evaluate the resultant linear BSDEs.
Its justification for the Lipschitz driver was recently given by Takahashi \& Yamada (2013)~\cite{TY}.
We refer to Fujii \& Takahashi (2012b)~\cite{FT-12b} for an example of explicit calculation,
and Fujii \& Takahashi (2014)~\cite{FT-14} as efficient Monte Carlo implementation where 
the analytical calculation is too cumbersome. 
See also Shiraya \& Takahashi (2014)~\cite{ST} and Cr\'epey \& Song (2014)~\cite{Crepey-Song}
 as concrete applications of the proposed perturbation method to the so-called credit valuation adjustment (CVA).

In this section, we propose a different type of perturbative expansion method.
In contrast to the method \cite{FT-12a}, it does not require the perturbation of the driver
and allows simpler analysis.
The new method directly expands the BSDE around the small-diffusion limit of 
the associated forward SDE, which only yields 
a system of linear ODEs to be solved at each order of expansion.

\subsection{A perturbative expansion scheme}
Let us first explain the idea of our perturbation scheme.
We will provide the justification and error estimate later.
For clarity of demonstration, let us assume a single security case.
The extension to a multiple security case is straightforward.
We introduce the underlying factor process $X:\Omega\times[t,T]\rightarrow \mbb{R}^d$ with 
an arbitrary starting time $t\in[0,T]$,  which follows
the SDE (do not confuse it with the position size process):
\bea
X_s^{t,x}=x+\int_t^s\mu(u, X^{t,x}_u)du+\int_t^s\sigma(u,X^{t,x}_u)dW_u~.
\eea
where $\mu:[t,T]\times\mbb{R}^d\rightarrow \mbb{R}^d$ and $\sigma:[t,T]\times \mbb{R}^d\rightarrow \mbb{R}^{d\times d}$.
The superscript $(t,x)$ indicates the initial condition for the process,
which will be omitted if it is clear from the context.
Let a function $f:\mbb{R}^d\times\mbb{R}\rightarrow \mbb{R}$ be defined by
\bea
f(x,v):=-\left(\frac{1}{M(x)}+\frac{\lambda(x)}{v+\eta(x)}\right)v^2+\gamma(x)
\eea
where $\xi,M,\gamma, \eta,\lambda:\mbb{R}^d\rightarrow \mbb{R}$, and consider the BSDE
\bea
V_t^{t,x}=\xi(X_T^{t,x})+\int_t^T f(X_s^{t,x},V_s^{t,x})ds-\int_t^T Z_s^{t,x} dW_s
\eea
where $V:\Omega\times[t,T]\rightarrow \mbb{R}$, $Z:\Omega\times[t,T]\rightarrow \mbb{R}^{d\times d}$.
This BSDE corresponds to (\ref{eq-v2}) with stochastic coefficients driven by a Markovian 
factor process $X$.
\subsubsection*{Assumption $P$}
{\it
1. The coefficients $\mu,\sigma$ are bounded Borel functions, 
and $\mu(t,x)$ and $\sigma(t,x)$ are continuous in (t,x) and smooth in x 
with bounded derivatives of all orders.\\
2. There exist constants $a_1,a_2>0$ such that for $\forall y\in\mbb{R}^d$
and $\forall (t,x)\in[0,T]\times\mbb{R}^d$, 
\be
a_1|y|^2\leq y^\top [\sigma\sigma^\top](t,x)y\leq a_2|y|^2.
\ee
3. $\xi(x),M(x),\gamma(x),\eta(x),\lambda(x)$ are bounded smooth functions of $x\in\mbb{R}^d$ 
and satisfy Assumptions $A$ and $B$ with $\forall x\in\mbb{R}^d$.
They are also assumed to have bounded derivatives of all orders.
}

Now, in order to approximate this process, we introduce a small parameter $\ep\in]0,1]$ and $\ep$-dependent process $X^\ep$
\bea
dX_s^{t,x, \ep}=\mu(s,X_s^{t,x,\ep})ds+\ep\sigma(s,X_s^{t,x,\ep})dW_s\quad s\in[t,T], \quad X_t^{t,x,\ep}=x~.
\eea
By using this process, we carry out small-diffusion asymptotic expansion of the system.
The associated perturbed BSDE is given by
\bea
V_t^{t,x,\ep}=\xi(X_T^{t,x,\ep})+\int_t^T f(X_s^{t,x,\ep},V_s^{t,x,\ep})ds-\int_t^T Z_s^{t,x,\ep} dW_s~.
\eea
At the moment, let us assume the differentiability in terms of $\ep$ 
so that we have
\bea
&&X_s^{t,x,\ep}=X_s^{[0]}+\ep X_s^{[1]}+\ep^2 X_s^{[2]}+\cdots~ \nn \\
&&V_s^{t,x,\ep}=V_s^{[0]}+\ep V_s^{[1]}+\ep^2 V_s^{[2]}+\cdots~ \nn \\
&&Z_s^{t,x,\ep}=Z_s^{[0]}+\ep Z_s^{[1]}+\ep^2 Z_s^{[2]}+\cdots~
\eea
for $s\in[t,T]$, where we have defined
\bea
V_s^{[n]}:=\frac{1}{n!}\frac{\part^n}{\part \ep^n}V_s^{t,x,\ep}\Bigr|_{\ep=0}, \quad s\in[t,T]
\eea
and similarly for the others.

For the zero-th order, $X^{[0]}$ and $V^{[0]}$ are given by the solutions of the ODEs:
\bea
&&\frac{dX_s^{[0]}}{ds}=\mu(s,X_s^{[0]})\quad s\in[t,T], \quad X_t^{[0]}=x \\
&&\frac{dV_s^{[0]}}{ds}=-f(X_s^{[0]},V_s^{[0]}) \quad s\in[t,T], \quad V_T^{[0]}=\xi(X_T^{[0]})
\eea
which corresponds to a deterministic case discussed in the previous section.
Thanks to Assumption $P$,  the above Riccati ODE has a positive bounded solution. 
We obviously have $Z^{[0]}=0$.

In the first order,  we have the linear FBSDE system:
\bea
&&dX_s^{[1]}=\part_x \mu^{0}(s) X_s^{[1]}ds+\sigma^0(s)dW_s~~s\in[t,T], \quad X_t^{[1]}=0 \\
&&V_t^{[1]}=\part_x \xi^{0}(T) X_T^{[1]} +\int_t^T\left\{\LDis \part_v f^{0}(s)V_s^{[1]}+\part_x f^0(s) X_s^{[1]}\right\}
ds-\int_t^T Z_s^{[1]}dW_s \nn \\
\eea
where we have used the short hand notation:
\bea
\mu^{0}(s):=\mu(s,X_s^{[0]}),~~\sigma^{0}(s):=\sigma(s,X_s^{[0]}),~~\xi^{0}(T):=\xi(X_T^{[0]}),~~f^{0}(s):=f(X_s^{[0]},V_s^{[0]})~.
\eea
which are all deterministic functions. We have also used $\part_x:=(\part/\part x_i)_{1\leq i\leq d}$, $\part_v:=\part/\part v$.
By the assumptions we made, it is clear that there exists a unique solution for the BSDE
satisfying $V^{[1]}\in\mbb{S}^p(t,T)$, $Z^{[1]}\in\mbb{H}^p(t,T)$ for $\forall p>1$.
In fact, it is straightforward to explicitly solve it as 
\bea
V_s^{[1]}=y(s)^\top X_s^{[1]}, \quad s\in[t,T]
\eea
where $y:[t,T]\rightarrow \mbb{R}^d$ is the solution of the linear ODE:
\bea
&&\frac{d [y(s)]_i}{ds}=-\sum_{j=1}^d\part_{x_i}\mu^{0}_j(s)[y(s)]_j-\part_v f^{0}(s)[y(s)]_i-\part_{x_i}f^{0}(s)
\quad s\in[t,T]\nn \\
&&y(T)=\part_x\xi^0(T).
\eea
Due to the assumption on boundedness, $Z^{[1]}=y(s)^\top \sigma^0(s)$ is actually uniformly bounded.

In the second order expansion, one can find
\bea
&&dX_s^{[2]}=\Bigl(\part_x\mu^0(s)X_s^{[2]}+\frac{1}{2}(X_s^{[1]})^\top\bigl(\part_x^2 \mu^0(s)\bigr)X_s^{[1]}\Bigr)ds
+X_s^{[1]}\part_x\sigma^0(s)dW_s,~~s\in[t,T]\nn \\
&&X_t^{[2]}=0
\eea
and
\bea
&&V_t^{[2]}=\part_x\xi^0(T)X_T^{[2]}+\frac{1}{2}(X_T^{[1]})^\top\bigl(\part_x^2\xi^0(T)\bigr)X_T^{[1]}+\int_t^T\Bigl\{ \part_v f^{0}(s)V_s^{[2]}+\part_x f^{0}(s)X_s^{[2]} \nn \\
&&+\frac{1}{2}\part_v^2 f^0(s)(V_s^{[1]})^2+\part_{x,v}f^{0}(s)X_s^{[1]}V_s^{[1]}+
\frac{1}{2}(X_s^{[1]})^\top \bigl(\part_x^2 f^{0}(s)\bigr)X_s^{[1]}\Bigr\}ds-\int_t^T Z_s^{[2]}dW_s,\nn\\
\label{eq-v2nd}
\eea
which is also a linear BSDE.
In this case, one has the solution of the following from:
\be
V^{[2]}_s=y_2(s)^\top X_s^{[2]}+(X_s^{[1]})^\top y_1(s)X_s^{[1]}+y_0(s), \quad s\in[t,T]. 
\label{eq-hypo-v2nd}
\ee
Here, $y_2:[t,T]\rightarrow \mbb{R}^d$, $y_1:[t,T]\rightarrow \mbb{R}^{d\times d}$ and $y_0:[t,T]\rightarrow \mbb{R}$ 
are defined as the solution 
of the next linear ODE system for $s\in[t,T]$, $i,j\in\{1,\cdots,d\}$:
\bea
&&\frac{d[y_2(s)]_i}{ds}=-\Bigl(\sum_{j=1}^d\part_{x_i} (\mu^{0}(s))_j[y_2(s)]_j+\part_v f^0(s)[y_2(s)]_i\Bigr)-\part_{x_i}f^0(s)\nn \\
&&\frac{d[y_1(s)]_{i,j}}{ds}=-\Bigl(\sum_{k=1}^d\Bigl[[y_1(s)]_{i,k}\part_{x_j}(\mu^0(s))_k
+[y_1(s)]_{j,k}\part_{x_i}(\mu^0(s))_k\Bigr]+\part_v f^0(s)[y_1(s)]_{i,j}\Bigr)\nn\\
&&\qquad\qquad -\frac{1}{2}\sum_{k=1}^d[y_2(s)]_k\part_{x_i,x_j}(\mu^0(s))_k
-\frac{1}{2}\part_v^2f^0(s)[y(s)]_i[y(s)]_j\nn \\
&&\qquad\qquad -\frac{1}{2}\Bigl([y(s)]_j\part_{x_i}+[y(s)]_i\part_{x_j}\Bigr)\part_vf^{0}(s)-\frac{1}{2}\part_{x_i,x_j}f^0(s)\nn \\
&&\frac{dy_0(s)}{ds}=-\part_v f^0(s)y_0(s)-{\rm Tr}\bigl[y_1(s)\sigma^0(s)(\sigma^0(s))^\top\bigr]
\eea
with the terminal conditions:
\be
y_2(T)=\part_x \xi^0(T),\quad [y_1(T)]_{i,j}=\frac{1}{2}\part_{x_i,x_j} \xi^0(T),\quad y_0(T)=0~.
\ee
The above results can be obtained by applying It\^o formula to (\ref{eq-hypo-v2nd})
and comparing its drift to that of (\ref{eq-v2nd}).
See Fujii~(2015)~\cite{Fujii} as a related idea of expansion of BSDEs.

These procedures can be repeated to an arbitrary higher order.
In each order, one can show that there exists a unique solution 
$V^{[n]}\in\mbb{S}^p(t,T)$, $Z^{[n]}\in \mbb{H}^p(t,T)$ for $\forall p>1$ due to the 
boundedness assumptions and the linearity of the BSDE (See Theorem 5.17 in \cite{Pardoux}.).
The solution can be expressed by some polynomials 
of $\{X^{[i]}\}_{1\leq i\leq n}$ with the coefficients to be determined by the linear ODE system.
Note that, for every order $n$, $X^{[n]}\in\mbb{S}^p(t,T)$ for $\forall p>1$.

\subsection{Convergence}
\begin{theorem}
Under Assumptions $A$, $B$ and $P$,
there exists some positive constants $C,C^\prime$, which are independent of $\ep$,  such that
\bea
\mbb{E}\Bigl|\Bigl|V^{\ep}-\Bigl(V^{[0]}+\sum_{n=1}^N \ep^n V^{[n]}\Bigr)\Bigr|\Bigr|_{[t,T]}^p
&\leq& \ep^{p(N+1)}C,\\
\mbb{E}\left(\int_t^T \Bigl|Z_s^{\ep}-\Bigl(Z_s^{[0]}+\sum_{n=1}^N \ep^n Z^{[n]}\Bigr)\Bigr|^2ds\right)^{p/2} 
&\leq&\ep^{p(N+1)}C^\prime
\eea 
for $\forall p>1$ and every positive integer $N$.
\begin{proof}
The arguments for the justification and convergence are similar to those of \cite{TY} 
and we sketch them in the following~\footnote{The author is grateful to prof. Takahashi for 
helpful discussions.}.
Under Assumption $P$, the continuity and differentiability of $X^{t,x,\ep}$ with respect to 
$\ep$ are well known (See, for example, \cite{T}.).
For the continuity and differentiability of $V^{\ep},Z^\ep$, we can follow the 
same arguments of Section 2.4 of El Karoui et.al.~(1997)~\cite{ElK} and
Theorem 3.1 of Ma \& Zhang (2002)~\cite{MaZhang}.
Their results are based on the popular estimate (See Lemma 2.2 of \cite{MaZhang})
for a BSDE with a Lipschitz driver.
Although the driver $f$ is not Lipschitz in our case, we can in fact use the 
same estimate. This is because, we know the solution $V^{\ep}$ is uniformly bounded and non-negative
thanks to Proposition \ref{P-v2-bound}.
For higher derivatives $(\part_\ep^k V^\ep, \part_\ep^k Z^\ep)_{k\geq 1}$, the arguments  are more straightforward since the BSDEs
are linear for them. One can check that the Lipschitz conditions are satisfied 
by the similar reasons. Thus, recursively,  the arguments in \cite{ElK,MaZhang} guarantee the continuity and differentiability 
up to an arbitrary order.

Now consider the $n$-th order derivatives
\bea
V_{n,s}^{t,x,\ep}:=\frac{\part^n}{\part \ep^n}V_s^{t,x,\ep}, \quad
Z_{n,s}^{t,x,\ep}:=\frac{\part^n}{\part \ep^n}Z_s^{t,x,\ep}, \quad
X_{n,s}^{t,x,\ep}:=\frac{\part^n}{\part \ep^n}X_s^{t,x,\ep}
\eea
and their restriction to the condition $\ep=0$
\bea
V^{[n]}_s=\frac{1}{n!}V_{n,s}^{t,x,\ep}\Bigl|_{\ep=0}~,\quad 
Z^{[n]}_s=\frac{1}{n!}Z_{n,s}^{t,x,\ep}\Bigl|_{\ep=0}~,\quad
X^{[n]}_s=\frac{1}{n!}X_{n,s}^{t,x,\ep}\Bigl|_{\ep=0}.
\eea
Taylor expansion gives  the associated BSDE as (omitting superscript (t,x)), for $s\in[t,T]$,
\bea
\label{eq-Vn-ep}
V_{n,s}^{\ep}=G_n+\int_s^T\Bigl(
H_{n,r}+\part_v f(X_r^\ep,V_r^\ep)V_{n,s}^\ep+\part_x f(X_r^\ep,V_r^\ep)X_{n,r}^\ep\Bigr)dr
-\int_s^T Z_{n,r}^\ep dW_r
\eea
where 
\bea
&&G_n=n!\sum_{k=1}^n \sum_{\beta_1+\cdots+\beta_k=n, \beta_i\geq 1}\frac{1}{k!}\part_x^k \xi(X_T^\ep)
\prod_{j=1}^k\frac{1}{\beta_j!}X_{\beta_j,T}^\ep 
\eea
and similarly
\bea
&&H_{n,r}=n!\sum_{k=2}^n \sum_{\beta_1+\cdots+\beta_k=n,\beta_i\geq 1}
\sum_{i=0}^{k}\sum_{j=k-i+1}^{k-i}\frac{1}{i!(k-i)!}\part_x^{k-i}\part_v^i f(X_r^\ep,V_r^\ep)\nn \\
&&\qquad \times \prod_{j=1}^{k-i}\frac{1}{\beta_j !}X_{\beta_j,r}^\ep
\prod_{l=k-i+1}^k\frac{1}{\beta_l!}V_{\beta_l,r}^\ep~.
\eea

One obtains the SDE for $X_{n,s}^\ep$ in a similar manner. For every $n$, due to Assumption $P$ and
the linearity of the SDE, one can show that $X_{n,\cdot}^\ep\in\mbb{S}^p(t,T)$
for every $\forall p>1$ and $\ep\in]0,1]$.
Thus $\mbb{E}\Bigl(|G_n|^p\Bigr)<\infty$ holds for $\forall p>1$.
Then, since (\ref{eq-Vn-ep}) is a linear BSDE,  one can show recursively that
$\mbb{E}\Bigl(\int_t^T |H_{n,s}|ds\Bigr)^p<\infty$ for $\forall p>1$ and
that there exists a unique solution $V^{\ep}_{n,\cdot}\in\mbb{S}^p(t,T)$, $Z^{\ep}_{n,\cdot}\in\mbb{H}^p(t,T)$
for every $n$, $\forall p>1$ and $\forall \ep\in]0,1]$,  by applying Theorem 5.17 in \cite{Pardoux}.

Using Taylor formula, one sees
\bea
&&V_s^\ep=V_s^{[0]}+\sum_{n=1}^N\frac{\ep^n}{n!}\frac{\part^n}{\part \ep^n}V_s^\ep\Bigr|_{\ep=0}
+\ep^{N+1}\int_0^1 \frac{(1-u)^N}{N!}\frac{\part^{N+1}}{\part \nu^{N+1}}V_s^{\nu}\Bigr|_{\nu=\ep u} du \nn \\
&&=V_s^{[0]}+\sum_{n=1}^N \ep^n V_s^{[n]}+\frac{\ep^{N+1}}{N!}\int_0^1 (1-u)^N V_{N+1,s}^{u\ep}du.
\eea
Thus, there exists some constant $C$ such that
\bea
\mbb{E}\Bigl|\Bigl|V^{\ep}-\Bigl(V^{[0]}+\sum_{n=1}^N \ep^n V^{[n]}\Bigr)\Bigr|\Bigr|_{[t,T]}^p
&\leq& \ep^{p(N+1)}C\int_0^1 \mbb{E} ||V_{N+1,\cdot}^{u\ep}||^p_{[t,T]} du 
\eea
and similarly, 
\bea
\mbb{E}\left(\int_t^T \Bigl|Z_s^{\ep}-\Bigl(Z_s^{[0]}+\sum_{n=1}^N \ep^n Z^{[n]}\Bigr)\Bigr|^2ds\right)^{p/2} 
&\leq& \ep^{p(N+1)}C\mbb{E}\left(\int_t^T \Bigl[\int_0^1 |Z^{u\ep}_{N+1,s}|^2du\Bigr]ds\right)^{p/2}\nn 
\eea
for every $N$ and $\forall p>1$, which proves the claim.
\end{proof}
\end{theorem}
The above result can easily be extended to the multiple security case by using Assumptions $A^\prime$, $B^\prime$
and the boundedness of $V_2$ proved in Theorem~\ref{T-v2-multi}.
Once the terminal penalty is replaced by a random variable $\xi\in\mbb{L}^p(\Omega)$, 
the proposed perturbation algorithm can be applied to a different class of BSDEs~\cite{Jeanblanc}, too.
Detailed numerical tests and the extension to more general class of BSRDEs~\cite{Tang-2003}
will be left for an important future work.

\section{Concluding Remarks}
In this paper, we  discussed the optimal position management strategy
for a maker maker who faces uncertain in- and out-flow of customer orders.
The optimal strategy is represented by the solution of the stochastic Hamilton-Jacobi-Bellman 
equation which is decomposed into three (one non-linear and two linear) BSDEs.
We provided the verification of the solution using the standard BSDE techniques 
for the single-security case and an interesting connection to a special type of SLQC
problem for the multiple-security case.
We also proposed a perturbative approximation technique for the relevant BSRDE,
which only requires a system of linear ODEs to be solved at each order of expansion.
Its justification and error estimate were also given.

Assuming general $\mbb{F}$-adaptedness (instead of $\mbb{F}^W$-adaptedness) 
of the relevant parameters looks an interesting extension of the proposed framework. 
This situation will arise when one introduces 
simultaneous jumps in the parameters, such as $M$, and the executions in the dark pool.
In this case, the driver of the resultant BSRDE depends on the martingale coefficient
of the counting process. As long as we know, the existence and uniqueness of the solution 
for the corresponding BSRDE have not yet been
proved.

It looks also interesting to combine a stochastic filtering for the intensity of customer orders.
Introducing a hidden Markov process, for example,  is likely to help
to model possible herding behavior among the customer orders.
See a related work Fujii \& Takahashi (2015)~\cite{FT-15}
on the mean-variance hedging problem for fund and insurance managers.

\section*{Acknowledgement}
The author gratefully acknowledges the helpful comments and useful discussions with
Akihiko Takahashi.  The author is also grateful to Kenichiro Shiraya and Taiga Saito 
for useful discussions. This research is partially supported by Center for Advanced Research in Finance (CARF).

\begin{appendix}
\section{The Property of the Terminal Position Size}
In this appendix, we study the behavior of the remaining inventory $X_T^{\opi,\odel}$ 
at the terminal time
according to the change of the penalty size.
We shall prove that it can be made arbitrary small by 
increasing the size of the penalty $\xi$.
This result implies that the proposed strategy can be considered as 
a generalization of the optimal liquidation solution in the existing literature
to the situation with uncertain customer orders.

Let us take a positive constant $1<L<\infty$ and 
set $\xi=L$, i.e., $\wt{\xi}=\beta_T/2+L$.
We denote the corresponding solutions of the BSDEs (\ref{eq-v2}), (\ref{eq-v1}) and
(\ref{eq-v0}) by $(V_i^L,Z_i^L)_{\{i=1,2,3\}}$, respectively.

\subsubsection*{Assumption~C}
{\it Take the lower bound $c$ in the assumption $(b_3)$ in such a way that
$c/(1+\bar{\lambda})<1$ and also $\wt{c}:=c/[\bar{M}(1+\bar{\lambda})]<1/2$.
Obviously, one can always choose $c>0$ (or equivalently $\bar{M}, \bar{\lambda}$)
to satisfy these inequalities.}

\begin{lemma} Under Assumptions A, B, C and $\xi=L$, the following 
inequalities hold for every $0\leq t\leq s\leq T$ with an $L$-independent positive constant $C$; 
\bea
&&V_2^L(t)\leq C\frac{1}{T-t+\ep_L} \nn \\
&&\exp\Bigl(-\int_t^s r(u,V_2^L(u))du\Bigr)\leq 
\Bigl(\frac{T-s+\ep_L}{T-t+\ep_L}\Bigr)^{\wt{c}}
\eea
where $\displaystyle \ep_L:=\frac{1}{L}$, $\displaystyle \wt{c}:=\frac{c}{\bar{M}(1+\bar{\lambda})}$ and
$\displaystyle r(t,y):=\Bigl(\frac{1}{M_t}+\frac{\lambda_t}{y+\eta_t}\Bigr)y$.
\label{L-v2-pre}
\begin{proof}
Since the inequality in Proposition~\ref{P-v2-bound} holds arbitrary $\ep>0$, one can choose $\ep=\ep_L=1/L$.
Then one obtains
\bea
&&V_2^L(t) \leq \frac{\ep_L}{(T-t+\ep_L)^2}+
\frac{T-t}{(T-t+\ep_L)^2}\bar{M}+\frac{\bar{\gamma}}{3}
\Bigl( (T-t+\ep_L)-\frac{\ep_L^3}{(T-t+\ep_L)^2}\Bigr) \nn \\
&&\leq \frac{1}{T-t+\ep_L}\Bigl(1+\bar{M}+\frac{\bar{\gamma}}{3}(T+1)^2\Bigr)\nn \\
&&\leq \frac{C}{T-t+\ep_L}~.
\eea

Similarly, 
\bea
V_2^L(t)\geq \frac{1}{\displaystyle \mbb{E}\left[\frac{1}{L}+\int_t^T \Bigl(\frac{1}{M_s}+\frac{\lambda_s}{\eta_s}\Bigr)ds
\Bigr|\calf_t^W\right]}\geq
\frac{1}{\displaystyle \ep_L+\frac{(1+\bar{\lambda})}{c}(T-t)}
\nn\\
\eea
where $c>0$ is the lower bound given in $(b_3)$. Thus,
\bea
&&\int_t^s r(u,V_2^L(u))du \geq \int_t^s \frac{1}{\ep_L+\frac{1+\bar{\lambda}}{c}(T-u)}\frac{1}{\bar{M}}du\nn \\
&&=-\frac{c}{\bar{M}(1+\bar{\lambda})}\ln\Bigl(
\frac{\ep_L+\frac{1+\bar{\lambda}}{c}(T-s)}{\ep_L+\frac{1+\bar{\lambda}}{c}(T-t)}\Bigr)~.
\eea
It yields
\bea
\exp\Bigl(-\int_t^s r(u,V_2^L(u))du\Bigr)\leq \left(
\frac{ \frac{c}{1+\bar{\lambda}}\ep_L+T-s}{ \frac{c}{1+\bar{\lambda}}\ep_L+T-t}\right)^{\wt{c}}~.
\eea
Note that for every $0\leq t\leq s\leq T$,
$
\Bigl(\frac{x\ep_L+T-s}{x\ep_L+T-t}\Bigr)^{\wt{c}}
$
is a increasing function for $x\geq 0$. Thus, due to the arrangement of $c$, one obtains
\bea
\exp\Bigl(-\int_t^s r(u,V_2^L(u))du\Bigr)\leq \Bigl(\frac{T-s+\ep_L}{T-t+\ep_L}\Bigr)^{\wt{c}}~.
\eea
\end{proof}
\end{lemma}

We also have the following Lemma.
\begin{lemma}Under Assumptions A, B, C and $\xi=L$,
there exists an $L$-independent positive constant $C$ such that
\bea
&&\mbb{E}\Bigl[||V_1^L||^2_T+||X^{\opi,\odel}(0,x)||^2_T\Bigr]\leq C~.
\eea
\label{L-Lind-bound}
\begin{proof}
Let us put $A:\Omega\times [0,T]\rightarrow \mbb{R}$ and $\alpha:\Omega\times[0,T]\rightarrow \mbb{R}$ as
\bea
&&A_u:=\Bigl(\frac{1}{M_u}+\frac{\lambda_u}{V_2^L(u)+\eta_u}\Bigr)\frac{S_u}{2}-\frac{1}{2}\Theta_u
-b_u\Psi_u \\
&&\alpha_u:=\frac{1}{2}\Bigl(\beta_u b_u \Psi_u-l_u\Bigr)~.
\eea
Obviously, $A,\alpha\in\mbb{S}^4(0,T)\subset \mbb{S}^2(0,T)$, and whose $\mbb{S}^2(0,T)$-norms
can be dominated by $L$-independent constants.
It is straightforward to check that $V_1^L$ can be written as
\bea
V_1^L(t)=-\mbb{E}\left[\int_t^T e^{-\int_t^s r(u,V_2^L(u))du}\Bigl(V_2^L(s)A_s-\alpha_s\Bigr)ds\Bigr|\calf_t^W\right]~.
\eea
Thus, by Lemma~\ref{L-v2-pre}, it satisfies the following inequality for $\forall t\in[0,T]$:
\bea
&&|V_1^L(t)|\leq \mbb{E}\left[\Bigl|\int_t^T e^{-\int_t^s r(u,V_2^L(u))du}(V_2^L(s)A_s-\alpha_s)ds\Bigr|
\Bigr|\calf_t^W\right]\nn \\
&&\leq (T-t)\mbb{E}\Bigl[||\alpha||_T\bigr|\calf_t^W\Bigr]+
\mbb{E}\Bigl[||A||_T\bigr|\calf_t^W\Bigr]\left(\int_t^T \Bigl(\frac{T-s+\ep_L}{T-t+\ep_L}\Bigr)^{\wt{c}}
\frac{C}{T-s+\ep_L}ds\right)\nn \\
&&\leq (T-t)\mbb{E}\Bigl[||\alpha||_T\bigr|\calf_t^W\Bigr]+C\mbb{E}\Bigl[||A||_T\bigr|\calf_t^W\Bigr]
\frac{1}{\wt{c}}\Bigl(1-\Bigl[\frac{\ep_L}{T-t+\ep_L}\Bigr]^{\wt{c}}\Bigr) \nn \\
&&\leq C\mbb{E}\left[ \Bigl. ||\alpha||_T+||A||_T \bigr|\calf_t^W\right]~.
\eea
Notice that $\Bigl(m_t:= \mbb{E}\Bigl[||\alpha||_T+||A||_T\bigr|\calf_t^W\Bigr] \Bigr)_{t\in[0,T]}$
is a square integrable martingale. 
Thus, from Doob's maximum inequality, one has
\bea
&&\mbb{E}\Bigl[ ||V_1^L||_T^2\Bigr]\leq C\mbb{E}\left[\sup_{t\in[0,T]}\bigl|m_t\bigr|^2\right]
\leq 4C\mbb{E}\Bigl[\bigl|m_T\bigr|^2\Bigr] \nn \\
&&\leq C\mbb{E}\Bigl[||\alpha||_T^2+||A||_T^2\Bigr]
\eea
where the right-hand side can be dominated by an $L$-independent constant.

Now, let us define another process $G:\Omega\times[0,T]\rightarrow \mbb{R}$ as
\be
G_u=\Bigl(\frac{1}{M_u}+\frac{\lambda_u}{V_2^L(u)+\eta_u}\Bigr)
\Bigl(V_1^L(u)+\frac{S_u}{2}\Bigr)+\frac{\Theta_u}{2}-\Phi_u~
\ee
which satisfies $G\in \mbb{S}^4(0,T)\subset \mbb{S}^2(0,T)$ and its $\mbb{S}^2(0,T)$ norm can be 
dominated by an $L$-independent constant by the first part of the proof.
From (\ref{eq-opi}), (\ref{eq-odel}) and (\ref{eq-xopt}), it is easy to see that
\bea
&&X_t^*=e^{-\int_0^t r(u,V_2^L(u))du}x-\int_0^t e^{-\int_s^t r(u,V_2^L(u))du}G_s ds \nn \\
&&\quad+\int_0^t\int_K e^{-\int_s^t r(u,V_2^L(u))du}z\wt{\caln}(ds,dz)
+\int_0^t e^{-\int_s^t r(u,V_2^L(u))du}\del_s^* d\wt{H}_s 
\label{eq-xt}
\eea
holds for every $t\in[0,T]$ (We used  the notation $X_t^*:=X_t^{\opi,\odel}(0,x)$.).
Using the fact that $r(\cdot, V_2^L(\cdot))$ is a positive process and the BDG inequality, 
we have, with some $L$-independent constant $C$, 
\bea
&&\mbb{E}\Bigl[||X^*||^2_t\Bigr]\leq C\mbb{E}\left[ x^2+||G||^2_t+\int_0^t \int_K z^2 \caln(ds,dz)
+\int_0^t |\del_s^{*}|^2 dH_s\right]~\nn \\
&&\leq C\mbb{E}\Bigl[x^2+||G||^2_T+||\Phi_2||_T+||V_1^L||_T^2+||S||^2_T\Bigr]+
C\mbb{E}\Bigl[\int_0^t ||X^*||_s^2 ds\Bigr]~.
\eea
Let denote an $L$-independent constant dominating the first term by $C^\prime$.
Since we already know $X^*\in\mbb{S}^4(0,T)$, 
\bea
\mbb{E}\Bigl[||X^*||_t^2\Bigr]\leq C^\prime +C\int_0^t \mbb{E}\Bigl[||X^*||_s^2\Bigr]ds
\eea
and hence by the Gronwall lemma, 
\bea
\mbb{E}\Bigl[||X^*||^2_T\Bigr]\leq C^\prime e^{CT}~.
\eea
Combining the first part, the claims were proved.
\end{proof}
\end{lemma}

Then, we can establish the following result.
\begin{theorem}
Under Assumptions A, B, C and $\xi=L$, there exists an $L$-independent positive constant $C$ satisfying
\be
\mbb{E}\Bigl[\bigl|X_T^{\opi,\odel}(0,x)\bigr|^2\Bigr]\leq C \Bigl(\frac{\ep_L}{T+\ep_L}\Bigr)^{2\wt{c}}
\ee
and hence one can make the terminal position size arbitrarily small by taking
a large $L<\infty$ as the penalty.
\begin{proof}
From (\ref{eq-xt}) and Lemma~\ref{L-v2-pre}, we have
\bea
&&\mbb{E}\Bigl[\bigl|X_T^*\bigr|^2\Bigr]\leq C\mbb{E}\left[
x^2\Bigl(e^{-\int_0^T r(u,V_2^L(u))du}\Bigr)^2+||G||_T^2
\Bigl(\int_0^T e^{-\int_s^T r(u,V_2^L(u))du}ds\Bigr)^2 \right. \nn \\
&&\hspace{20mm}\left.+\int_0^T e^{-2\int_s^T r(u,V_2^L(u))du}\Bigl(\Phi_{2,s}+|\del_s^*|^2\Bigr)ds\right] \nn 
\eea
\bea
&&\leq Cx^2\Bigl(\frac{\ep_L}{T+\ep_L}\Bigr)^{2\wt{c}} \nn \\
&&+C\mbb{E}\Bigl[||X^*||_T^2+||V_1^L||^2_T+||G||^2_T+||S||^2_T+||\Phi_2||_T\Bigl]
\int_0^T \Bigl(\frac{\ep_L}{T-s+\ep_L}\Bigr)^{2\wt{c}}ds~.
\eea
Notice that the expectation in the second term is dominated by an $L$-independent constant 
by Lemma~\ref{L-Lind-bound}. Using the assumption $2\wt{c}< 1$, we have
\bea
&&\mbb{E}\Bigl[\bigl|X_T^*\bigr|^2\Bigr]\leq C
\left\{\Bigl(\frac{\ep_L}{T+\ep_L}\Bigr)^{2\wt{c}}+\frac{1}{1-2\wt{c}}(T+\ep_L)\Bigl(\frac{\ep_L}{T+\ep_L}\Bigr)^{2\wt{c}}
-\frac{\ep_L}{1-2\wt{c}}\right\}\nn \\
&&\leq C \Bigl(\frac{\ep_L}{T+\ep_L}\Bigr)^{2\wt{c}}
\eea
with some $L$-independent positive constant, and hence obtained the desired result.
\end{proof}
\end{theorem}

\subsubsection*{Remark}
Although we can discuss the limit of the singular terminal condition $L\rightarrow \infty$
as presented in \cite{Jeanblanc}, we can only apply their results to $V_2^L$.
For $V_1^L$, there appears a singular drift term which is expected to create a discontinuity 
at the terminal point.  This makes the detailed analysis difficult to carry out.
However, as the previous result shows, we can make the terminal position size
arbitrarily small by selecting large enough $L<\infty$ as the penalty.
Therefore, the proposed strategy can also be used 
as an effective liquidation  strategy
in the presence of incoming customer orders for a market maker.

Although it is natural, even in a multiple-security setup,  to imagine that one can make the terminal position size arbitrarily small
by increasing the size of the eigenvalues of $\xi$.
Although it is intuitively clear, it is difficult to prove since we do not have 
an explicit expression for the upper/lower bound of $V_2$ any more. 

Let us suppose, in the interval $[T-\ep, T]$ with some constant $\ep>0$, that
$M,\gamma,\xi,\beta$ can be diagonalized by the common {\it constant} orthogonal matrix $O$.
In addition, suppose the market maker stops accepting the customer orders and stops using the dark pool.
Then, by considering the securities in the base $O^\top S$ and the corresponding 
positions $O^\top \bX$, the market maker's problem can be decomposed into 
$n$ single security liquidation problems. In this case, $\hat{V_2}:=O^\top V_2 O$ becomes 
diagonal process in $[T-\ep,T]$ and $\hat{V_1}:=O^\top V_1$ interacts
with the only one corresponding element of $\hat{V_2}$. In this special situation,
it is clear that the position can be made arbitrary small by the corresponding optimal strategy
thanks to the arguments made in the single security case.

\end{appendix}



\end{document}

%% file: tymacro3.TEX

\newtheorem{definition}{Definition}[section]
\newtheorem{assumption}{$[$ A}[section]
\newtheorem{condition}{$[$ C}
\newtheorem{lemma}{Lemma}[section]
\newtheorem{proposition}{Proposition}[section]
\newtheorem{theorem}{Theorem}[section]
\newtheorem{remark}{Remark}[section]
\newtheorem{example}{Example}[section]
\newtheorem{corollary}{Corollary}[section]
\def\n{{\bf n}}
\def\A{{\bf A}}
\def\B{{\bf B}}
\def\C{{\bf C}}
\def\D{{\bf D}}
\def\E{{\bf E}}
\def\F{{\bf F}}
\def\G{{\bf G}}
\def\H{{\bf H}}
\def\I{{\bf I}}
\def\J{{\bf J}}
\def\K{{\bf K}}
\def\L{{\bf L}}
\def\M{{\bf M}}
\def\N{{\bf N}}
\def\O{{\bf O}}
\def\P{{\bf P}}
\def\Q{{\bf Q}}
\def\R{{\bf R}}
\def\S{{\bf S}}
\def\T{{\bf T}}
\def\U{{\bf U}}
\def\V{{\bf V}}
\def\W{{\bf W}}
\def\X{{\bf X}}
\def\Y{{\bf Y}}
\def\Z{{\bf Z}}
\def\cala{{\cal A}}
\def\calb{{\cal B}}
\def\calc{{\cal C}}
\def\cald{{\cal D}}
\def\cale{{\cal E}}
\def\calf{{\cal F}}
\def\calg{{\cal G}}
\def\calh{{\cal H}}
\def\cali{{\cal I}}
\def\calj{{\cal J}}
\def\calk{{\cal K}}
\def\call{{\cal L}}
\def\calm{{\cal M}}
\def\caln{{\cal N}}
\def\calo{{\cal O}}
\def\calp{{\cal P}}
\def\calq{{\cal Q}}
\def\calr{{\cal R}}
\def\cals{{\cal S}}
\def\calt{{\cal T}}
\def\calu{{\cal U}}
\def\calv{{\cal V}}
\def\calw{{\cal W}}
\def\calx{{\cal X}}
\def\caly{{\cal Y}}
\def\calz{{\cal Z}}
%
\def\sskip{\hspace{0.5cm}}
\def\simleq{ \raisebox{-.7ex}{\em $\stackrel{{\textstyle <}}{\sim}$} }
\def\leqsim{ \raisebox{-.7ex}{\em $\stackrel{{\textstyle <}}{\sim}$} }
\def\ep{\epsilon}
\def\half{\frac{1}{2}}
\def\iku{\rightarrow}
\def\Iku{\Rightarrow}
\def\ikup{\rightarrow^{p}}
\def\inclusion{\hookrightarrow}
\def\cadlag{c\`adl\`ag\ }
\def\up{\uparrow}
\def\down{\downarrow}
\def\doti{\Leftrightarrow}
\def\douti{\Leftrightarrow}
\def\dochi{\Leftrightarrow}
\def\douchi{\Leftrightarrow}%
\def\yy{\\ && \nonumber \\}
\def\y{\vspace*{3mm}\\}
\def\nn{\nonumber}
\def\be{\begin{equation}}
\def\ee{\end{equation}}
\def\bea{\begin{eqnarray}}
\def\eea{\end{eqnarray}}
\def\beas{\begin{eqnarray*}}
\def\eeas{\end{eqnarray*}}
%
\def\hd{\hat{D}}
\def\hv{\hat{V}}
\def\hsd{{\hat{d}}}
\def\hx{\hat{X}}
\def\hsx{\hat{x}}
\def\bsx{\bar{x}}
\def\bsd{{\bar{d}}}
\def\bx{\bar{X}}
\def\ba{\bar{A}}
\def\bb{\bar{B}}
\def\bc{\bar{C}}
\def\bv{\bar{V}}
\def\balpha{\bar{\alpha}}
\def\bbalpha{\bar{\bar{\alpha}}}
\def\combi{\l(\begin{array}{c}\alpha\\ \beta \end{array}\r)}
\def\f{^{(1)}}
\def\s{^{(2)}}
\def\ss{^{(2)*}}
\def\l{\left}
\def\r{\right}
\def\a{\alpha}
\def\b{\beta}
\def\L{\Lambda}